\documentclass[jgrga]{AGUTeX}

\usepackage{lineno}
\usepackage{array,multirow,makecell}
 \usepackage{color}

  \usepackage{graphicx}
  \setkeys{Gin}{draft=false}
\authorrunninghead{CHARNAY ET AL.}

\titlerunninghead{3D MODELLING OF THE CLIMATES OF THE ARCHEAN EARTH}

\begin{document}

%% ------------------------------------------------------------------------ %%
%
%  TITLE
%
%% ------------------------------------------------------------------------ %%

\title{EXPLORING THE FAINT YOUNG SUN PROBLEM AND THE POSSIBLE CLIMATES OF THE ARCHEAN EARTH WITH A 3D GCM}
%
% e.g., \title{Terrestrial ring current:
% Origin, formation, and decay $\alpha\beta\Gamma\Delta$}
%

%% ------------------------------------------------------------------------ %%
%
%  AUTHORS AND AFFILIATIONS
%
%% ------------------------------------------------------------------------ %%

%Use \author{\altaffilmark{}} and \altaffiltext{}

% \altaffilmark will produce footnote;
% matching \altaffiltext will appear at bottom of page.

\authors{B. Charnay\altaffilmark{1},
 F. Forget\altaffilmark{1}, R. Wordsworth\altaffilmark{2},
 J. Leconte\altaffilmark{1}, E. Millour \altaffilmark{1}, F. Codron \altaffilmark{1}, and A. Spiga \altaffilmark{1}}

\altaffiltext{1}{Laboratoire de M\'et\'eorologie Dynamique, Universit\'e P\&M Curie (UPMC), Paris, France.}
\altaffiltext{2}{Department of the Geophysical Sciences, University of Chicago, Chicago, Illinois, USA.}

%\altaffiltext{2}{Department of Geography, Ohio State University,
%Columbus, Ohio, USA.}

%\altaffiltext{3}{Department of Space Sciences, University of
%Michigan, Ann Arbor, Michigan, USA.}

%\altaffiltext{4}{Division of Hydrologic Sciences, Desert Research
%Institute, Reno, Nevada, USA.}

%\altaffiltext{5}{Dipartimento di Idraulica, Trasporti ed
%Infrastrutture Civili, Politecnico di Torino, Turin, Italy.}

%% ------------------------------------------------------------------------ %%
%
%  ABSTRACT
%
%% ------------------------------------------------------------------------ %%

% >> Do NOT include any \begin...\end commands within
% >> the body of the abstract.

\begin{abstract}

Different solutions have been proposed to solve the 'faint young Sun problem', defined by the fact that the Earth was not fully frozen during the Archean despite the fainter Sun.
Most previous studies were performed with simple 1D radiative convective models and did not account well for the clouds and ice-albedo feedback or the atmospheric and oceanic transport of energy. 
We apply a Global Climate Model (GCM) to test the different solutions to the faint young Sun problem. We explore the effect of greenhouse gases (CO$_2$ and CH$_4$), atmospheric pressure, cloud droplet size, land distribution  and Earth's rotation rate. We show that, neglecting organic haze, 100 mbars of CO$_2$ with 2 mbars of CH$_4$ at 3.8 Ga and 10 mbars of CO$_2$ with 2 mbars of CH$_4$ at 2.5 Ga allow a temperate climate (mean surface temperature between 10$^\circ$C and 20$^\circ$C). 
Such amounts of greenhouse gases remain consistent with the geological data.
Removing continents produces a warming lower than +4$^\circ$C. The effect of rotation rate is even more limited.
Larger droplets (radii of 17 $\mu m$ versus 12 $\mu m$) and a doubling of the atmospheric pressure produce a similar warming of around +7$^\circ$C. In our model, ice-free waterbelts can be maintained up to 25$^\circ$ N/S with less than 1 mbar of CO$_2$ and no methane. An interesting cloud feedback appears above cold oceans, stopping the glaciation. Such a resistance against full glaciation tends to strongly mitigate the faint young Sun problem.

\end{abstract}

%% ------------------------------------------------------------------------ %%
%
%  BEGIN ARTICLE
%
%% ------------------------------------------------------------------------ %%

% The body of the article must start with a \begin{article} command
%
% \end{article} must follow the references section, before the figures
%  and tables.

\begin{article}

%% ------------------------------------------------------------------------ %%
%
%  TEXT
%
%% ------------------------------------------------------------------------ %%

\section{Introduction and background}
\subsection{The faint young Sun problem}
The Archean is the geological era following the Hadean (starting with Earth formation 4.56 Ga ago) and preceding the Proterozoic. It starts at 3.8 Ga, after the Late Heavy Bombardment (LHB), and ends at 2.5 Ga with the Great Oxidation Event. The first reported fossils of bacteria date back 3.5 Ga ago \cite[]{schopf06} and there is possibly evidence for life from carbon isotopes up to 3.8 Ga \cite[]{mojzsis96,rosing99}. The emergence of life is believed to have occurred before 3.5 Ga, and maybe even before the LHB \cite[]{nisbet01}. Thus studying the climates of the Archean Earth is of prime interest to understand the environment in which life emerged and evolved. 
\\
According to the standard model of stellar evolution, the Sun was 20 to 25 $\%$ weaker during the Archean \cite[]{gough81}. With such a weaker Sun, the Earth with the present-day atmospheric composition would fall into a full glaciation, hardly reconcilable with the evidence of liquid water and life during the whole Archean \cite[]{sagan72, feulner12}. This has been named the 'faint young Sun problem' or the 'faint young Sun paradox'.

While the Earth was unfrozen during most of the Archean, there is geological evidence for glaciations at the end of the Archean (the Huronian glaciations), between 2.45 and 2.22 Ga \cite[]{evans97}. These have been linked to the rise of oxygen in the atmosphere \cite[]{kasting06a, kasting06b} and the destruction of atmospheric methane. There was also a possible glaciation at 2.9 Ga \cite[]{young98} but it would have been regional, not global, and its origin remains unknown \cite[]{kasting06b}. The Archean Earth therefore seems to have experienced very few glaciations, implying that there were temperate or hot climates during most of the Archean. However, the Archean rock record is extremely sparse and the latitudes of these geological data are unknown. Sea ice and continental ices could have existed at high latitudes, and other glaciations may have occured.

Paleotemperatures were estimated from the isotopic composition of marine cherts \cite[]{knauth03, robert06} and indicate hot oceans (between 60$^\circ$C and 80 $^\circ$C) during the Archean. This makes the faint young Sun problem even more challenging. The validity of these measurements has been questioned, however, because of the possible variation of the isotopic composition of oceans during the time \cite[]{kasting06a, kasting06c, jaffres07} and because of the impact of hot hydrothermal circulation on chert formation \cite[]{vandenboorn07}. The most recent analyzes obtained an upper limit at 40$^\circ$C \cite[]{hren09, blake10}. 

Estimating the temperature of the oceans is also controversial as far as biological evidence is concerned. Genetic evolution models suggest that the ancestors of bacteria, eukaryota and archaea were thermophylic during the Archean \cite[]{gaucher08,boussau08} consistent with the oceans at 60$^\circ$C to 80$^\circ$C. The environment of the last universal common ancestor (LUCA) has been estimated to be more temperate (around 20$^\circ$C), suggesting an adaptation to high temperatures \cite[]{boussau08} possibly in response either to a change in the climate of the early Earth, or to the strong impacts during the LHB. Yet these trends do not necessary correspond to the climate of the Archean Earth, but maybe just reflect the environment where life was thriving.
To avoid glaciation and allow such temperate or hot climates, the early Earth must have experienced warming processes.

\subsection{Solutions to the faint young Sun problem}
A different atmospheric composition with a larger amount of greenhouse gases was first proposed as the key to get a habitable Earth under a fainter Sun. The first studies examined ammonia (NH3) \cite[]{sagan72}, a strong greenhouse gas. However, ammonia would have had a short lifetime (less than 40 years) due to photolysis in the high atmosphere. It would therefore not have been present in sufficient amounts unless there was a large, permanent surface source \cite[]{kuhn79}. Yet, such a permanent source would have produced so much N$_2$ (compared to present inventory) by photolysis of ammonia that it could not happen.

Current thinking is that the early Earth had a CO$_2$ and CH$_4$ rich atmosphere. The amount of CO$_2$ in the atmosphere is controlled by the carbonate-silicate cycle \cite[]{walker81}, which acts as a thermostat on the climate, preserving Earth from a full glaciation by injecting CO$_2$ from volcanoes in the atmosphere. CO$_2$ may have reached large amounts during the early Earth, although the maximum is not well known (between 0.1 and 10 bars \cite[]{walker85,sleep01}). According to 1D models \cite[]{vonparis08, pavlov00}, with a 20 $\%$ weaker Sun, $\sim$0.03 bar of CO$_2$ is required to raise the temperature above the frost point, and $\sim$0.2 bar to get present-day temperatures. However, geochemical data from paleosols constrain the maximum partial pressure of CO$_2$ to around 0.02 bar for the end of the Archean \cite[]{rye95, sheldon06, driese11, feulner12}, 10 times lower to what is required to get a temperate climate. A stronger constraint of 0.9 mbar of CO$_2$ has been obtained \cite[]{rosing10} based on the coexistence of siderite and magnetite in archean banded iron formations. However, this contradicts other measurements \cite[]{hessler04,sheldon06,driese11}  and is currently debated \cite[]{reinhard11}. 

Methane has been suggested as an important complement to CO$_2$ to warm the early Earth\cite[]{kiehl87}. It can absorb thermal radiation at 7-8$\mu$m, thus at the edge of the atmospheric window (8-12 $\mu$m), where CO$_2$ cannot. It can therefore produce an efficient warming. In an anoxic atmosphere, the lifetime of methane is 1000 times higher than today \cite[]{zahnle86, kasting06a}. During the Archean, methane would have been released by methanogenic bacteria through the reaction:
\begin{equation} 
CO_2 + 4H_2 \Rightarrow CH_4 +2H_2O
\end{equation} 
where H$_2$ comes from hydrothermal sources and volcanoes, or from the primitive atmosphere \cite[]{tian05}.

With the present day biological flux, the archean atmosphere would contain around 3 mbar of methane \cite[]{kasting06a}. Based on the biological flux and the escape rate of hydrogen, the amount of methane is estimated to be of the order of 1 mbar during the Archean, with a plausible range between 0.1 mbar and 35 mbar \cite[]{kharecha05}. However, the time when methanogens appeared and diversified is still highly uncertain. Attempts to determine this time have been made using genomic evolution models \cite[]{battistuzzi04,house03}
but the resulting times range from the beginning to the end of the Archean.

Before methanogens appeared or when they were confined to hydrothermal vents, methane was present in the atmosphere but in lower amounts. \cite{tian11} estimate around 0.5-5 $\times$ 10$^{-3}$ of methane for the prebiotic atmosphere, based on emanations from the Lost City hydrothermal vent field studied by \cite{kelley05}.

If the mixing ratio of methane is large, an organic haze forms. This is expected to happen when the CH$_4$/CO$_2$ ratio becomes higher than 0.1-0.3 according to photochemical models and experimental data \cite[]{zerkle12, trainer06}. In addition to limiting the amount of methane, the formation of haze could produce an anti-greenhouse effect \cite[]{mckay91, haqq-misra08, kasting06b}, and hence cool the Earth. The impact of this anti-greenhouse effect on surface temperature is unknown, mostly because the fractal nature of haze particles is unconstrained. Fractal particles produce a limited anti-greenhouse effect, compared to spherical particles. Moreover, they act as a UV shield, like the ozone layer, protecting both life and photolytically unstable reduced gases \cite[]{wolf10}. Under this shielding, ammonia may have been maintained in sufficient amount to solve the faint young Sun problem.

In any event, measurements of carbon isotopes appear consistent with a methane-rich atmosphere at the end of the Archean with possible episodic formation of haze \cite[]{zerkle12}. 
According to the 1D model of \cite{haqq-misra08}, which includes haze formation, 1 mbar of methane in an atmosphere containing 20 mbar of CO$_2$ allows to reach present-day temperatures at the end of the Archean. However, an ice-free Earth cannot be maintained with only a CO$_2$-CH$_4$ greenhouse warming consistent with the geological constraints for CO$_2$.

Given the difficulties in reconciling the warm temperatures estimated for the Archean with geological constraints for CO$_2$, other mechanisms of warming than greenhouse gases have been explored. Clouds both warm the surface by absorbing and reemitting infrared radiation and cool it by reflecting sunlight in the visible.
Through these mechanisms, lower clouds tend to globally cool the Earth, while higher clouds tend to warm it. A negative feedback, increasing the amount of cirrus (higher clouds), was considered to keep the climate clement: the "Iris hypothesis" \cite[]{lindzen01, rondanelli10} but remains controversial \cite[]{lin02, goldblatt11}.
A more plausible hypothesis is that lower clouds were optically thinner during the Archean, owing to the lack of cloud condensation nuclei from biological sources, which yields a decrease of the planetary albedo, and hence a warming ($\sim$+10$^\circ$C) \cite[]{rosing10}. 

The planetary albedo has been suggested to be lower in the Archean because of the reduced surface of emerged continents \cite[]{rosing10}. 
It has also been proposed that the pressure was higher in the past, because the equivalent of around 2 bars of nitrogen is present in the Earth's mantle \cite[]{goldblatt09}. That nitrogen, initially in the atmosphere, should have been incorporated by subduction (probably by biological fixation). Therefore, it is plausible that the partial pressure of nitrogen reached 2 to 3 bars during the Archean. According to 1D modelling, doubling the amount of present-day atmospheric nitrogen  would cause a warming of 4-5 $^\circ$C \cite[]{goldblatt09}. Besides, hydrogen could have been abundant in the early Earth's atmosphere. The lack of O$_2$ would have led to a cooler exosphere limiting the hydrogen escape. Thus, the balance between hydrogen escape and volcanic outgassing could have maintained a hydrogen mixing ratio of more than 30 $\%$ \cite[]{tian05}.

The combination of an hydrogen-rich atmosphere with a higher atmospheric pressure (2 to 3 bars) would produce an important greenhouse effect by collision absorption of H$_2$-N$_2$, sufficient to get present-day temperatures with a limited amount of CO$_2$ \cite[]{wordsworth13}. 
These mechanisms remain to be further explored. They might not be sufficient to solve the faint young Sun problem alone, although they probably played a role in maintaining a clement climate, complementing the greenhouse effect by CO$_2$ and methane.

\subsection{Previous modelling studies}

Given the paucity of available data for the early Earth, climate modelling is particularly useful to explore and understand the evolution of the atmosphere and the climate. 1D radiative-convective models allow different hypotheses to solve the faint young Sun problem to be tested \cite[]{owen79, kasting84, kiehl87,
kasting86, haqq-misra08, vonparis08, domagal-goldman08, goldblatt09, rosing10}. However, such models calculate the mean surface temperature below a single atmospheric column with averaged solar flux. Clouds are often omitted, or widely fixed (altitude, optical depth and covering). Furthermore, the transport of energy by the atmosphere and the ocean is not taken into account in 1D modelling. The continental and oceanic ice formation is not accounted for either. Thus, 1D radiative-convective models fail to capture both cloud and ice-albedo feedbacks, and transport processes, which are fundamental to determine the climate sensitivity under different conditions. Moreover, the lack of clouds in 1D radiative-convective models can lead to overestimates of the radiative forcing of greenhouse gases \cite[]{goldblatt11}.
\\
The most accurate way to simulate the climate is to use 3D Global Climate Models (GCM), including more of the fundamental processes which control climate sensitivity (e.g. clouds, oceanic transport, continental and oceanic sea ice). Studies of the Archean Earth using GCMs are rare. However, preliminary GCM studies showed that the absence of an ozone layer, continent, and a faster rotation rate could modify cloud coverage and hence the surface temperature \cite[]{jenkins93a, jenkins93b, jenkins95,jenkins99}. 
Using a 3D oceanic model coupled to a parameterized atmospheric model, \cite{kienert13} explored the key role of the ice-albedo feedback and found that 0.4 bar of CO$_2$ is required to avoid full glaciation. This illustrates the key role of ice-albedo feedback.

We describe below the application of a new generic GCM recently developed in our team to Archean climates. The versatility of this model allowed us to explore the climates of the Archean Earth under many conditions discussed in the literature (such as greenhouse gases, atmospheric pressure and rotation rate).  Our goal is to test different warming processes suggested by 1D models to better constraint the Archean climate and adress key questions left unresolved by 1D models.

\section{Description of the model}
\subsection{Generalities}

We use a new ”generic” version of the LMD Global Climate Model recently developed to simulate a complete range of planetary atmospheres. This model has been used to study early climates in the solar system \cite[]{forget12,wordsworth12} as well as climates on extrasolar planets \cite[]{wordsworth11, leconte13}. The model is derived from the LMDZ Earth GCM \cite[]{hourdin06}, which solves the primitive equations of meteorology using a finite difference dynamical core on an Arakawa C grid. This dynamical core has been in a variety of atmospheres such as the present Earth \cite[]{hourdin06}, Mars \cite[]{forget99},  Venus \cite[]{lebonnois10}, and Titan \cite[]{lebonnois12, charnay12}. 

In this paper, simulations were performed with a horizontal resolution of 64x48 (corresponding to resolutions of 3.75$^\circ$ latitude by 5.625$^\circ$ longitude). In the vertical, the model uses hybrid coordinates, that is, a terrain-following $\sigma$ coordinate system ($\sigma$ is pressure divided by surface pressure) in the lower atmosphere, and pressure levels in the upper atmosphere. In this work, we used 20 layers, with the lowest mid-layer level at 4 m and the top level at 3 hPa ($>$ 50 km). Nonlinear interactions between explicitly resolved scales and subgrid-scale processes are parameterized by applying a scale-selective horizontal dissipation operator based on an n time iterated Laplacian $\Delta^n$ . This can be written as 
$\partial q/\partial t = ([-1]^n /\tau_{diss})(\delta x)^{2n} \Delta^n q $
where $\delta x$ is the smallest horizontal distance represented in the model and $\tau diss$ is the dissipation timescale for a structure of scale $\delta x$. Subgrid-scale dynamical processes (e.g. turbulent mixing and convection) are parameterized as in \cite[]{forget99}. In practice, the boundary layer dynamics are accounted for by \cite{mellor82} unstationary 2.5-level closure scheme, plus a “convective adjustment” which rapidly mixes the atmosphere in the case of dry unstable temperature profiles. Turbulence and convection mix energy (potential temperature), momentum (wind), and water vapor. A standard roughness coefficient of $z_0$ = 10$^{-2}$ m is used for both rocky and ocean surfaces for simplicity.
The evolution of surface temperature is governed by the balance between radiative fluxes (direct solar insolation, and thermal radiation from the atmosphere and the surface), turbulent and latent heat fluxes, and thermal conduction in the soil. The parameterization of this last process is based on an 18-layer soil model solving the heat diffusion equation using finite differences. The depth of the layers were chosen to capture diurnal thermal waves as well as the deeper annual thermal wave. A vertically homogeneous soil is assumed. The thermal inertia for ground is set to 2000 J s$^{-1/2}$ m$^{-2}$ K$^{-1}$ everywhere. 

In equilibrium, the globally averaged difference between OLR and ASR was found to be lower than 2 W/m$^2$. Simulations were typically run for 40 years for temperate climates. For cold climates, the equilibrium is reached after a longer time. We ran simulations for 80 years to be sure not to miss a full glaciation.

\subsection{Radiative transfer}

Our radiative scheme is based on the correlated-k model as in \cite{wordsworth11}, with absorption for H$_2$O, CO$_2$ and CH$_4$ calculated directly from high resolution spectra obtained by a line-by-line model that uses of the HITRAN 2008 database \cite[]{rothman09}. At a given pressure and temperature, correlated-k coefficients in the GCM are interpolated from a matrix of coefficients stored in a 7 x 9 temperature and log-pressure grid: T = 100, 150, 200, 250, 300, 350, 400 K, p = 10$^{−1}$ , 10$^0$ , 10$^1$ , ...., 10$^7$ Pa. We used 36 spectral bands in the thermal infrared and 38 at solar wavelengths. Sixteen points were used for the g-space integration, where g is the cumulated distribution function of the absorption data for each band. 

The radiative transfer code used the two stream scheme from \cite{toon89}.
Rayleigh scattering by N$_2$ and CO$_2$ molecules was included using the method described in \cite{hansen74}.
The water vapor continuum from \cite{clough92} is included. This is important essentially for hot climates. 
An improved collision induced absorption (CIA) for CO$_2$ is used \cite{wordsworth10}. This effect is important in the case of a high CO$_2$ partial pressure, but remains small for the simulation presented in this paper. The sublorentzian profiles of \cite{perrin89} are used for the CO$_2$ far line absorption.

The model predicts the cloud cover at each grid mesh. For each column, a global cloud cover is fixed equal to the cover of the optically thicker cloud. This allows not to be biased by thin stratospheric clouds, more present without ozone. Then, we assume that each individual cloud cover is equal to this global cover with a maximum overlap.
Radiative transfer is therefore computed twice: in a clear sky column, and in a cloudy sky column whose area is equal to the global cloud cover.

\subsection{Water cycle}

\subsubsection{Cloud formation}
Cloud formation is computed through a moist convective scheme \cite[]{manabe67} and a large-scale condensation scheme \cite[]{letreut91}. We have chosen the Manabe and Wetherald scheme rather than the Betts-Miller scheme because it is more robust for a wide range of pressure, at the cost of giving enhanced precipitation at the equator \cite[]{frierson07}. The large-scale condensation scheme predicts a cloud cover for every cell in the atmosphere. The latent heat between icy and liquid phase for cloud is not taken into account . Supercooling is taken into account, allowing liquid droplet down to -18 $^\circ$C. The percentage of liquid water in the condensed phase is given by:

 \begin{equation}
R_{liq/condensed}=\frac{T-(273.15-18)}{18}
 \end{equation}

with T the temperature in K. For temperature higher (lower) than 273.15 K (255.15 K), clouds are only composed of liquid (ice) droplets.

The radius of cloud droplet is fixed (12 $\mu$m for liquid droplet and 35 $\mu$m for icy droplet for modern Earth).

\subsubsection{Precipitation}
Water precipitation is divided into rainfall and snowfall. We consider for both cases that precipitation is instantaneous (i.e. it goes directly to the surface) but can evaporate while falling through subsaturated layers. 
\\ 
Rainfall is parameterized using the scheme from \cite{boucher95}. Conversion of cloud liquid droplets to raindrops occurs by coalescence with other droplets.
The variations of cloud liquid water mixing ratio is given by:
\begin{equation}
\frac{dq_l}{dt}=-c \alpha \rho_{air}q_l^{2}r_l
\label{precip}
\end{equation}
where $c$ is a coefficient smaller than unity (we take $c=0.6$ to match the observations on present-day Earth),  $\alpha$=1.3x10$^5$ m$^2$kg$^{-1}$s$^{-1}$ , $\rho_{air}$ is the air volumic mass and r$_l$ is the cloud droplet radius.
\\ 
Snowfall is calculated using the falling velocity $v_i$ of icy particles. The variation of the cloud ice water mixing ratio is given by
\begin{equation}
\frac{dq_i}{dt}=-\frac{q}{\Delta z} v_i
\end{equation}
where $\Delta z$ is the depth of the layer in the model. Thus, we assume that what exits in a layer, falls to the ground instantaneously.
$v_i$ is taken equal to the terminal velocity for ice crystals \cite[]{langleben54, sekhon70}:
 \begin{equation}
v_i=10.6 \times r_i^{0.31}
 \end{equation}
with $v_i$ in m/s and $r_i$, the ice crystal radius, in m. We chose this parameterization fine-tuned for the terrestrial case rather than a more general Stokes sedimentation scheme because the particular shape of snowflakes are better accounted for. This parameterization is only valid for the Earth with a background mean molar mass pretty similar to the present day, which is the case in our simulations.
\\ 
For evaporation of rain and snow during the fall in the atmosphere, we use the following formula \cite[]{gregory95}:

\begin{equation} 
e=2. \times 10^{-5}(1-RH)\sqrt{P} 
\end{equation} 
where $e$ is the evaporation rate (in kg/m$^3$/s), $RH$ is the relative humidity and $P$ the precipitation flux (in kg/m$^2$/s). 

\subsubsection{Hydrology and evaporation}
The ground is modelled as a simple bucket model with a maximum water capacity of 150 kg/m$^2$. When the water quantity exceeds this limit, the surplus is regarded as runoff and added to the ocean. Snow/ice over ground are treated the same way, though with maximal limit of 3000 kg/m$^2$. The melting (formation) of snow/ice are computed with latent heat exchange when the surface temperature is lower (higher) than 273.15 K.

The ground albedo $A$ rises with the quantity of snow/ice up to a maximum value ($A_{max}$=0.55) as 
        \begin{eqnarray} 
A=A_{initial}+(A_{max}-A_{initial})\times q_{ice}/q_{*} 
    \end{eqnarray} 
with $q_{ice}$ the amount of snow/ice on the surface and $q_{*}$=33 kg/m$^2$ (corresponding to a uniform layer of approximately 3.5 cm thickness \cite[]{letreut91}).

Evaporation $E$ is computed within the boundary layer scheme, using a bulk aerodynamic formula multiplied by a dryness coefficient $\beta$:
 \begin{equation}
E=\beta \rho_{air} C_d V (a_s-a_a)
 \end{equation}

 \begin{equation}
\beta=min(1,2\frac{q}{150})
 \end{equation}
 \begin{equation}
C_d=\frac{\kappa}{ln(1+z_1/z_0)}
 \end{equation}
with $V$ the wind speed at the first level, $\rho$ the air density$a_s$ and $a_a$ the absolute humidity at the surface and at the first level, $q$ the mass of water (liquid + solid) at the surface (in kg/m$2$), $\kappa$ the constant of Von Karman (=0.4), $z_0$ the roughness and $z_1$ the altitude of the first level. $\beta$=1 over the ocean.

\subsection{Oceanic transport and sea ice}
The ocean and the sea ice are important components of the Earth's climate.  A
full ocean general circulation model would however be too expensive for our study, because of the long integrations required for complete
adjustment to the many different conditions imposed (changing luminosity,
continental extent, atmospheric composition, and so on).  Moreover such
precision is not required given the large number of unknown parameters. We
therefore use a simplified ocean model with fast adjustment, concentrating on a
representation of the essential components for the global climate:
the oceanic heat transport and the sea ice extent. 

We use the ocean model from \cite{codron12}, which uses the same horizontal grid
as the GCM. This model is composed of two layers. The first layer (50 m depth), represents the surface mixed layer, where the exchanges
with the atmosphere take place. The second layer (150 m depth) represents
the deep ocean.
The transport of heat by the ocean circulation is given by two components.
First, the impact of sub-grid scale eddies is represented by horizontal
diffusion, with a uniform diffusivity in both layers.  Then, the mean
wind-driven circulation is computed by calculating the Ekman mass fluxes in the
surface layer from the surface wind stress, and taking an opposite return flow
at depth. These mass fluxes are then used to advect the ocean temperature
horizontally. In the case of divergent horizontal mass fluxes, the upwelling or
downwelling mass flux is obtained by continuity. Other components of the ocean
circulation -- density-driven circulations and horizontal gyres -- are not
present in the model. Although they can play an important role regionally on
the present-Earth, they are weaker on global average, and gyres would be absent in
the case of a global ocean. This simplified model reproduces the
global meridional oceanic heat transport quite closely compared to a full GCM, both for
actual Earth and for a simulated global ocean case \cite[]{marshall07}. 

The oceanic model also computes the formation of oceanic ice. Sea ice forms
when the ocean temperature falls below -1.8$^\circ$C, and melts when its temperature rises above freezing. The changes in ice extent and thickness are computed based on energy conservation, keeping the ocean
temperature at -1.8$^\circ$C as long as ice is present.  A layer of snow can be present
above the ice. The surface albedo is then that of snow, or for bare ice
\begin{eqnarray} 
A=A_{ice}^{max}-(A_{ice}^{max}-A_{ice}^{min})\exp(-h_{ice}/h_{ice}^0)
\end{eqnarray} 
with $A$ the albedo, $A_{ice}^{max}$=0.65 the maximal albedo, $A_{ice}^{min}$=0.2 the minimal albedo, $h_{ice}$ the ice thickness (in m) and $h_{ice}^0$= 0.5 m. The albedo over the ice-free ocean is taken to be equal to 0.07.  
The value for the maximal sea ice albedo we used (i.e. 0.65) is classical for GCMs. It is a pretty high value for studies of snowball Earth \cite[]{abbot11}, making our results concerning cold climates pretty robust.

The transport of sea ice is not taken into account. This has a small impact for the present-day conditions, but it may be more important for different conditions (e.g. a colder climate with more sea ice) \cite[]{lewis03}.

\section{Model verification on modern Earth}
\subsection{Simulation of the modern Earth}
Before starting to investigate Earth's early climates, it is important to check the model's performance by simulating the present Earth. For verification, we compare the simulations we get with NCEP/NCAR reanalysis (www.esrl.noaa.gov/psd/data).

We use 360x180 resolution maps for topography and ground albedo which are interpolated on the GCM grid.
The average surface pressure is 1013 mbars, so a little higher than the reality because of the topography. Yet, the impact of this small change is negligible. 

The atmosphere is composed of N$_2$ with CO$_2$ and CH$_4$ at present-day levels (0.375 mbar of CO$_2$ and 1.75$\times$10$^{-3}$ mbar of CH$_4$). We do not include O$_2$ and ozone for simplicity in the radiative transfert calculation. The absence of O$_2$ has no strong impact (changes in Rayleigh scattering and heat capacity are small). The absence of ozone implies the absence of a temperature inversion in the stratosphere. The consequences of this assumption are discussed briefly in the next section. 
As far as astronomical parameters are concerned, we use the present-day obliquity (23.44$^\circ$), eccentricity (0.0167) and solar constant (1366 W/m$^2$).

\begin{figure*}
\centering
\noindent\includegraphics[width=20pc]{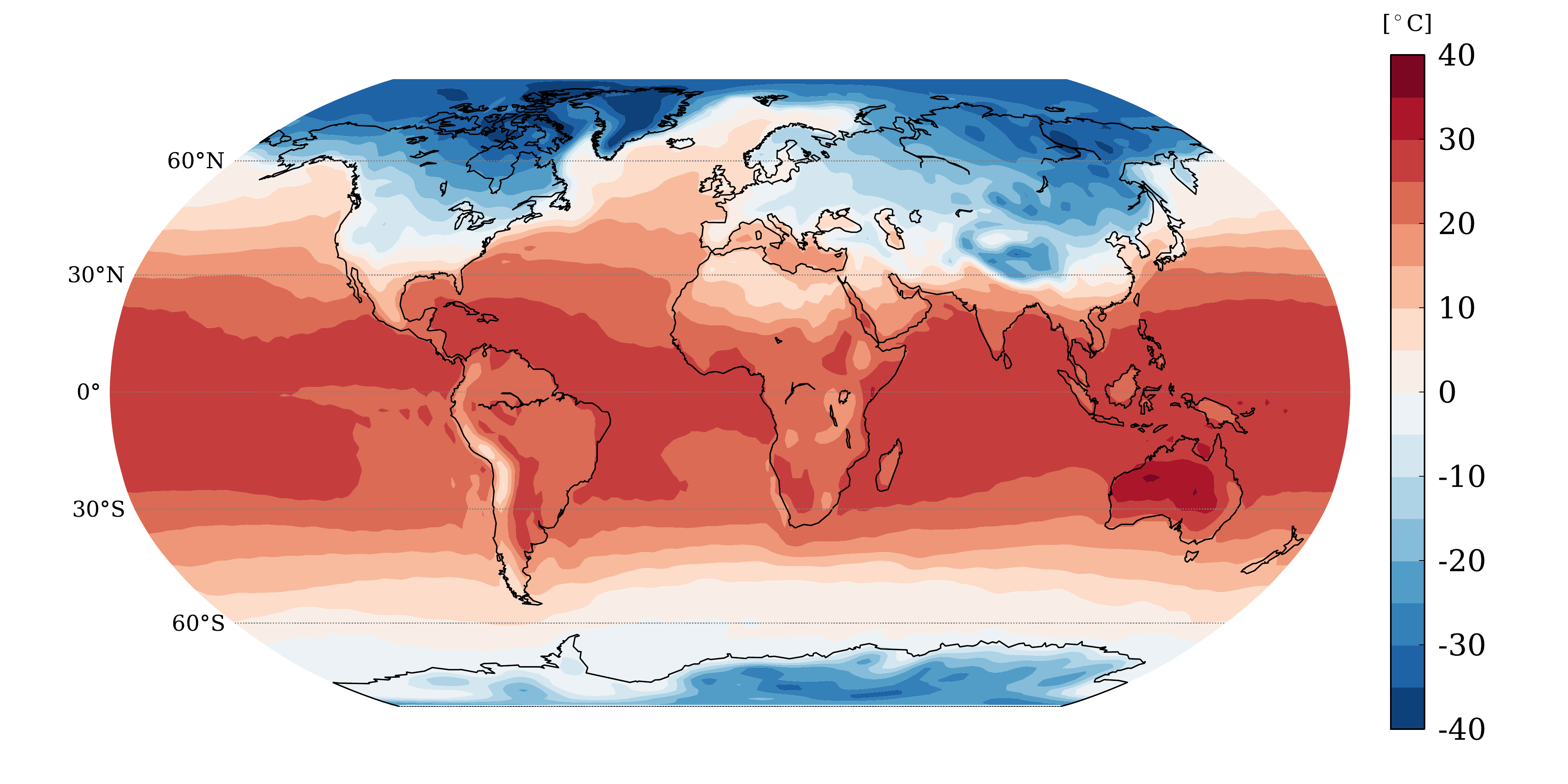}
\noindent\includegraphics[width=20pc]{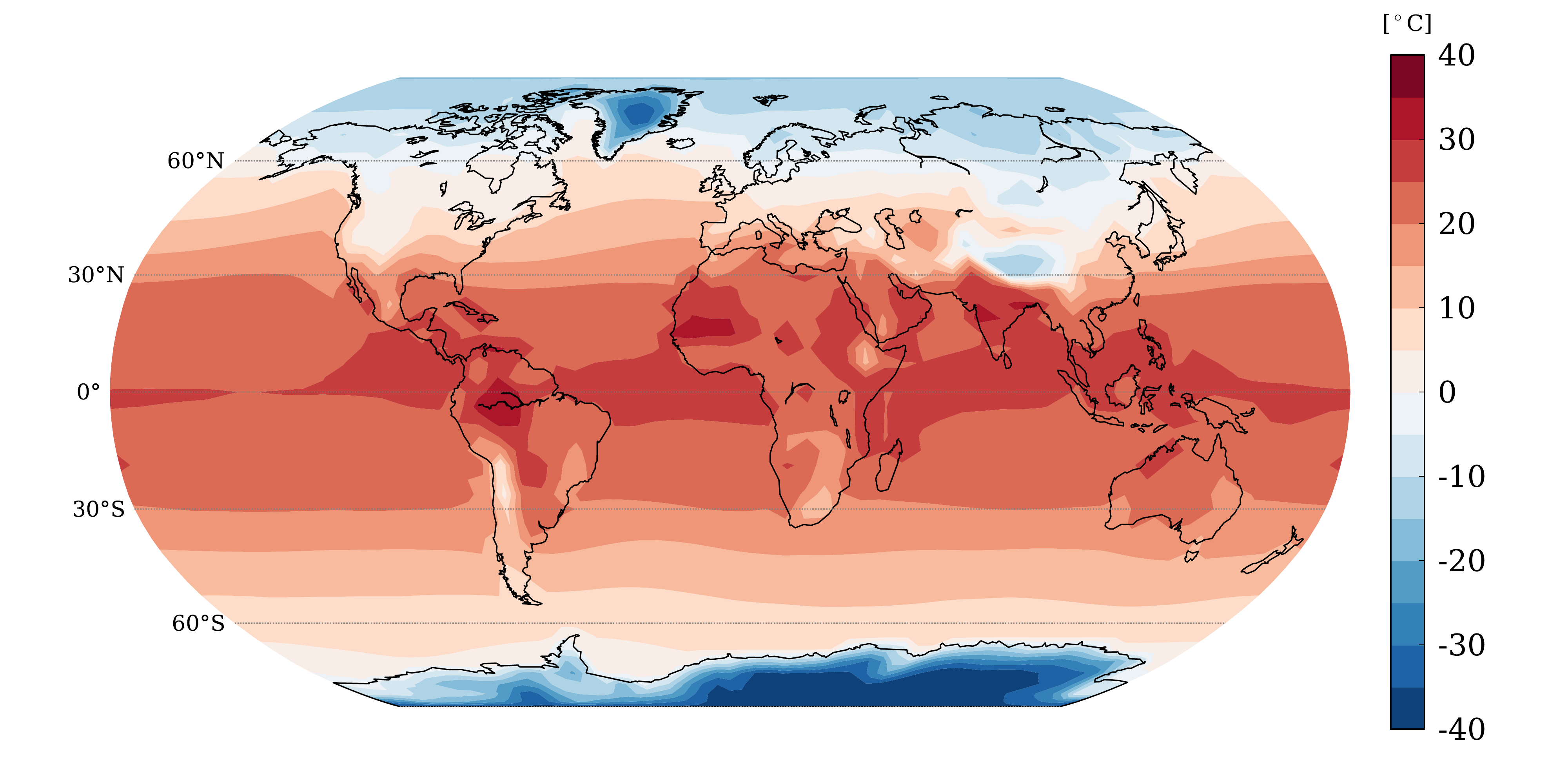}
\noindent\includegraphics[width=20pc]{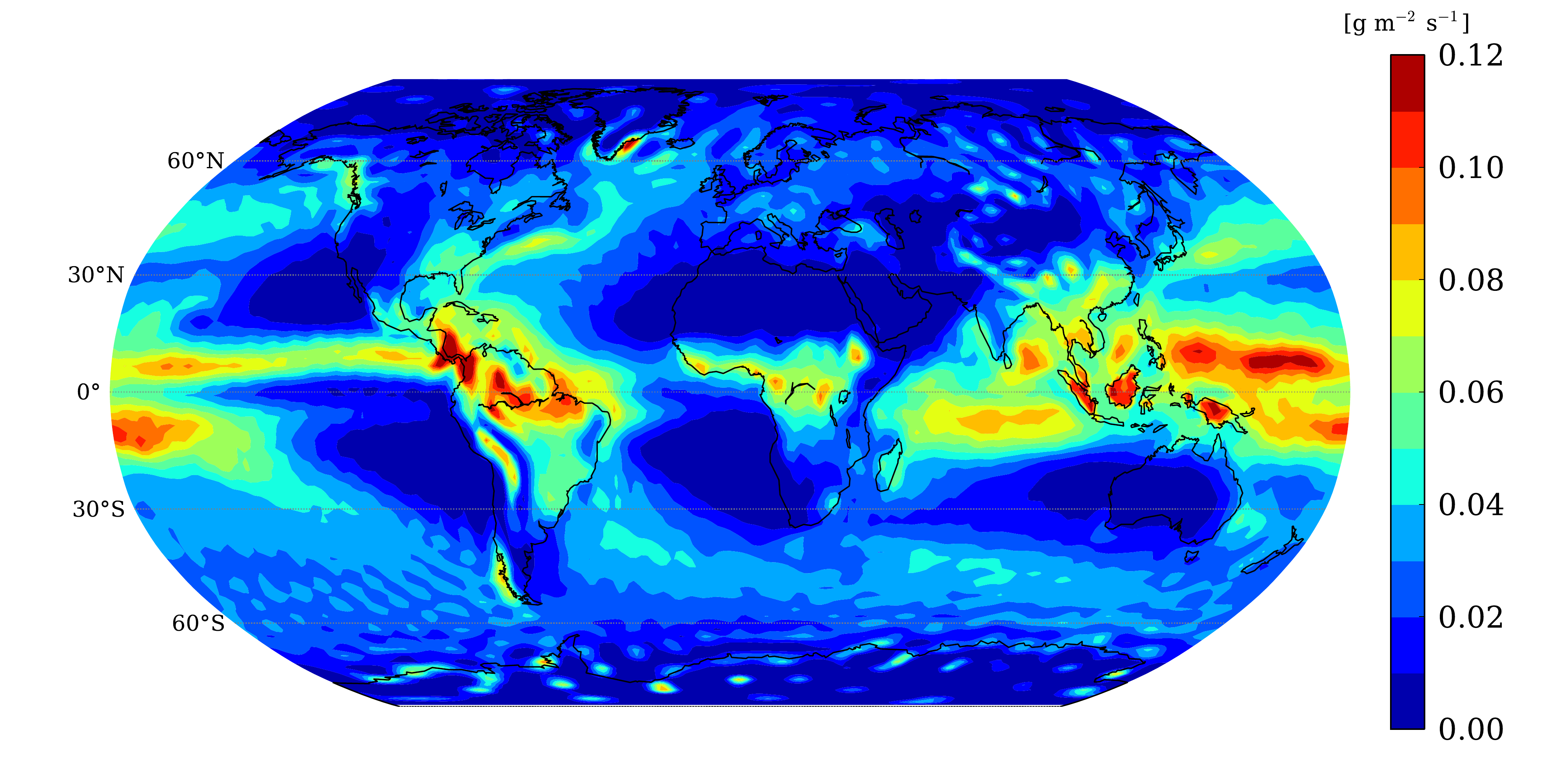}
\noindent\includegraphics[width=20pc]{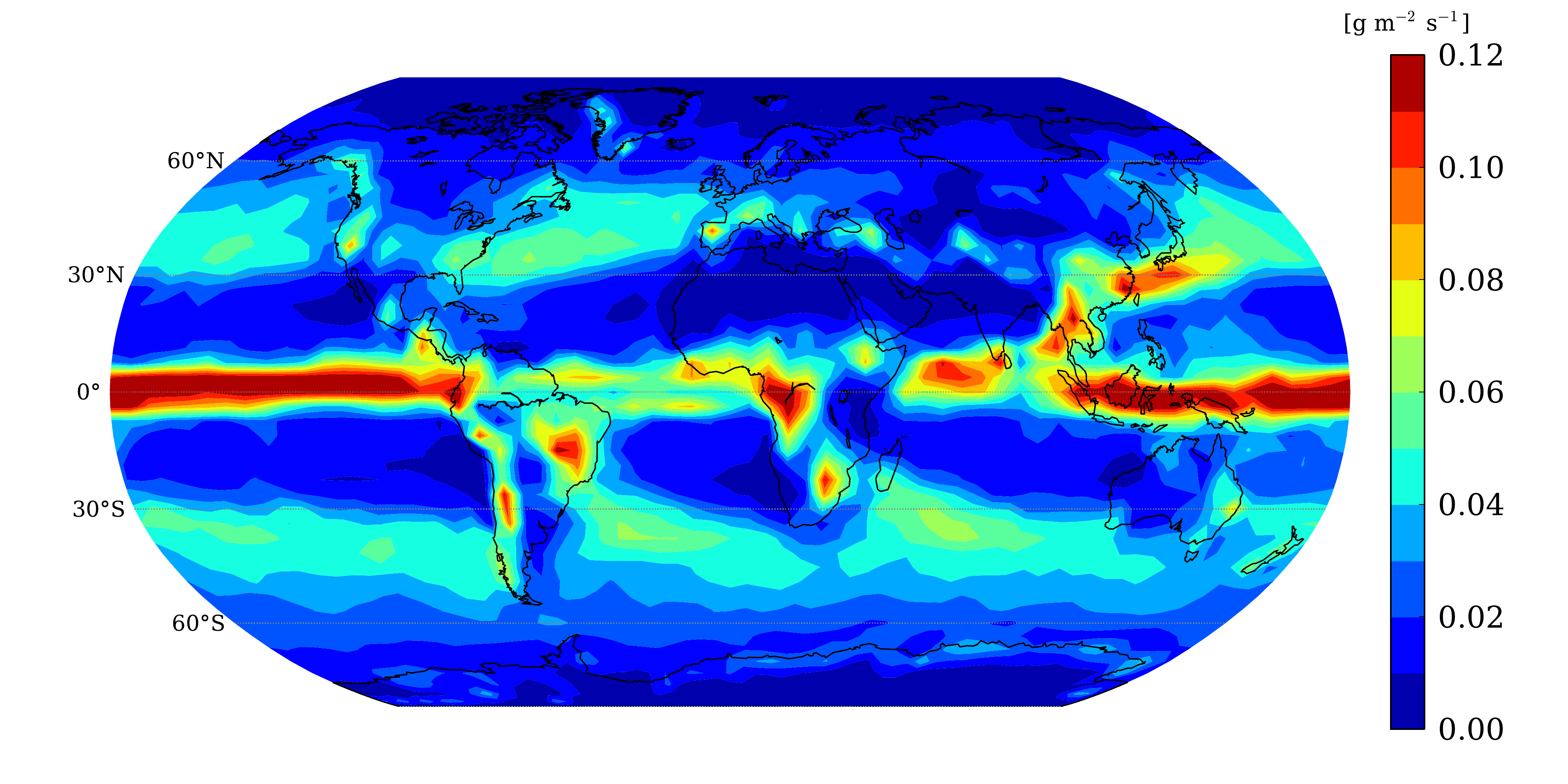}
\caption{Annually averaged surface temperature (top) and precipitation (bottom) on the Earth from NCEP/NCAR reanalysis (left) and from our GCM (right).}
\label{figure1}
\end{figure*}

In our Earth-like simulation, we find an average surface temperature of 14.9$^\circ$C (compared to 15.0$^\circ$C in NCEP reanalyses) and an average planetary albedo of 0.36 (compared to 0.33). This higher albedo in our model is due to a higher amount of clouds produced at the Inter-Tropical Convergence Zone (ITCZ) and the absence of ozone, which decreases solar atmospheric absorption and increases the higher clouds (see next paragraph). Since higher clouds are enhanced, their greenhouse effect increases and compensates the albedo cooling. Thus the surface temperature is not much affected. 
The top panel in figure \ref{figure1} shows the annual average surface temperature simulated by our model and from the NCEP/NCAR reanalysis. The tropical seas are a little colder (0.8$^\circ$C colder at the equator and 1.9$^\circ$C over all the tropics) in our model and the northern polar regions are warmer (15$^\circ$C warmer). Therefore, our GCM produces higher temperatures at high latitudes, maybe due to a too strong meridional transport or a cloud effect. In the northern polar region, this is strongly amplified by sea ice which disappears in summer.  If the sea ice was still present in summer, the temperatures would be far lower as in the southern hemisphere. Our model also produces a colder western Pacific warm pool. The difference of sea surface temperature is around 2-3$^\circ$C between western and eastern Pacific at the equator, thus less than the $\sim$5$^\circ$C observed. 

The bottom panel in figure \ref{figure1} shows the annual average precipitation simulated by our model and from the NCEP/NCAR reanalysis. Our model produces too much precipitation over the ITCZ in Pacific Ocean. This is certainly due to our moist convection scheme, as explained in \cite{frierson07}. The precipitation over the tropics and mid-latitudes is accurate with rain over Amazonia and very little rain over the Saharan and Arabian deserts.

To conclude, the temperatures and precipitations are globally close to the NCEP/NCAR reanalysis. Even if our model does not reproduce perfectly the latitudinal gradient and regional temperatures, it simulates the terrestrial climate with sufficient accuracy for the goals of this study (i.e. to simulate global climates, 3D effects and good trends for paleo-climates). Indeed, having prognostic equations for the cloud properties and distribution, the predictive hability of 3D models is enhanced when compared to 1D models, even when looking only at the global scale.

\subsection{Effect of the ozone and the oceanic transport}
The present ozone layer absorbs around 7 W/m$^2$ of the solar flux at UV wavelengths, producing the stratospheric thermal inversion. This absorption tends to cool the surface but it is compensated by the greenhouse effect of the ozone. For radiative-convective models, the presence of ozone yields a warming of the surface \cite[]{francois88}. The presence of ozone impacts the tropospheric dynamics \cite[]{kiehl88}, and also leads to a warmer higher troposphere, which could result in a decrease of the amount of cirrus clouds, reducing their greenhouse effect \cite[]{jenkins95}.
If ozone is added, both effects (the increase of solar atmospheric absorption and the decrease of amount of higher clouds) should decrease the planetary albedo without changing surface temperature very much. This should partially explain the bias for the albedo in our model.

Other 3D simulations with and without ozone will be required to estimate precisely the impact of ozone.

\begin{figure*}
\centering
\noindent\includegraphics[width=20pc]{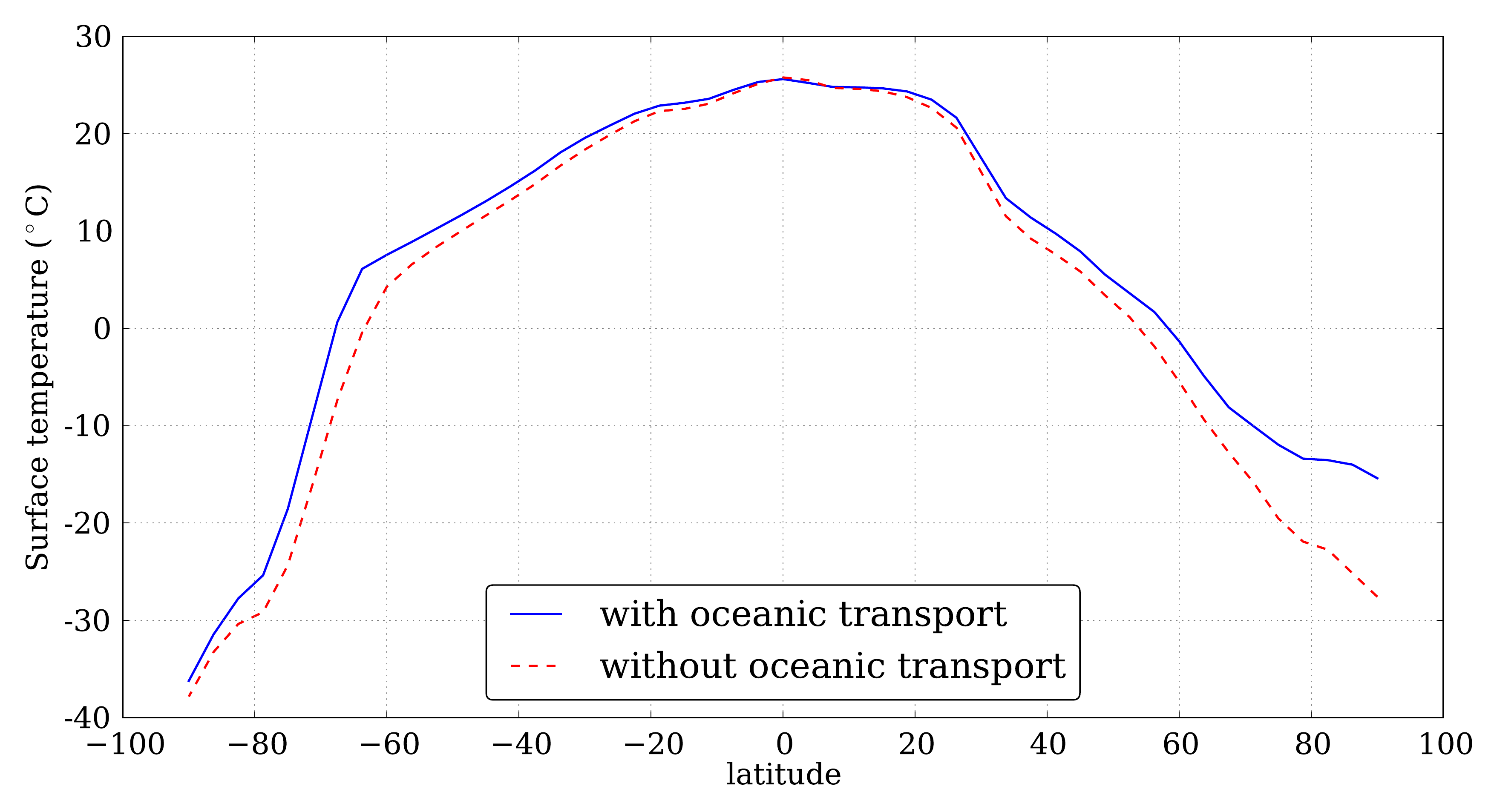}
\caption{Zonnaly averaged surface temperature on the modern Earth with (blue) and without (red) oceanic transport.}
\label{figure2}
\end{figure*}

The heat transport by the ocean can impact the mean global surface temperature. Under present-day conditions, the oceanic transport limits the spreading of sea ice and thus the ice-albedo feedback. This leads to a global warming of the Earth.
Figure \ref{figure2} shows the effect of the oceanic transport for Earth. Without oceanic transport, sea ice cover rises, yielding a decrease of $\sim$2$^\circ$C for the mean surface temperature and a decrease of 12$^\circ$C at the northern pole. Thus, the oceanic transport has a moderate impact in our model.

\section{Simulation of the Archean Earth}
In this section, we apply our model to the Archean Earth under a fainter Sun. We explore the effects of land distribution and greenhouse gases over the whole duration of the Archean.

\subsection{Testing the faint young Sun problem}
\begin{figure*}
\centering
\noindent\includegraphics[width=20pc]{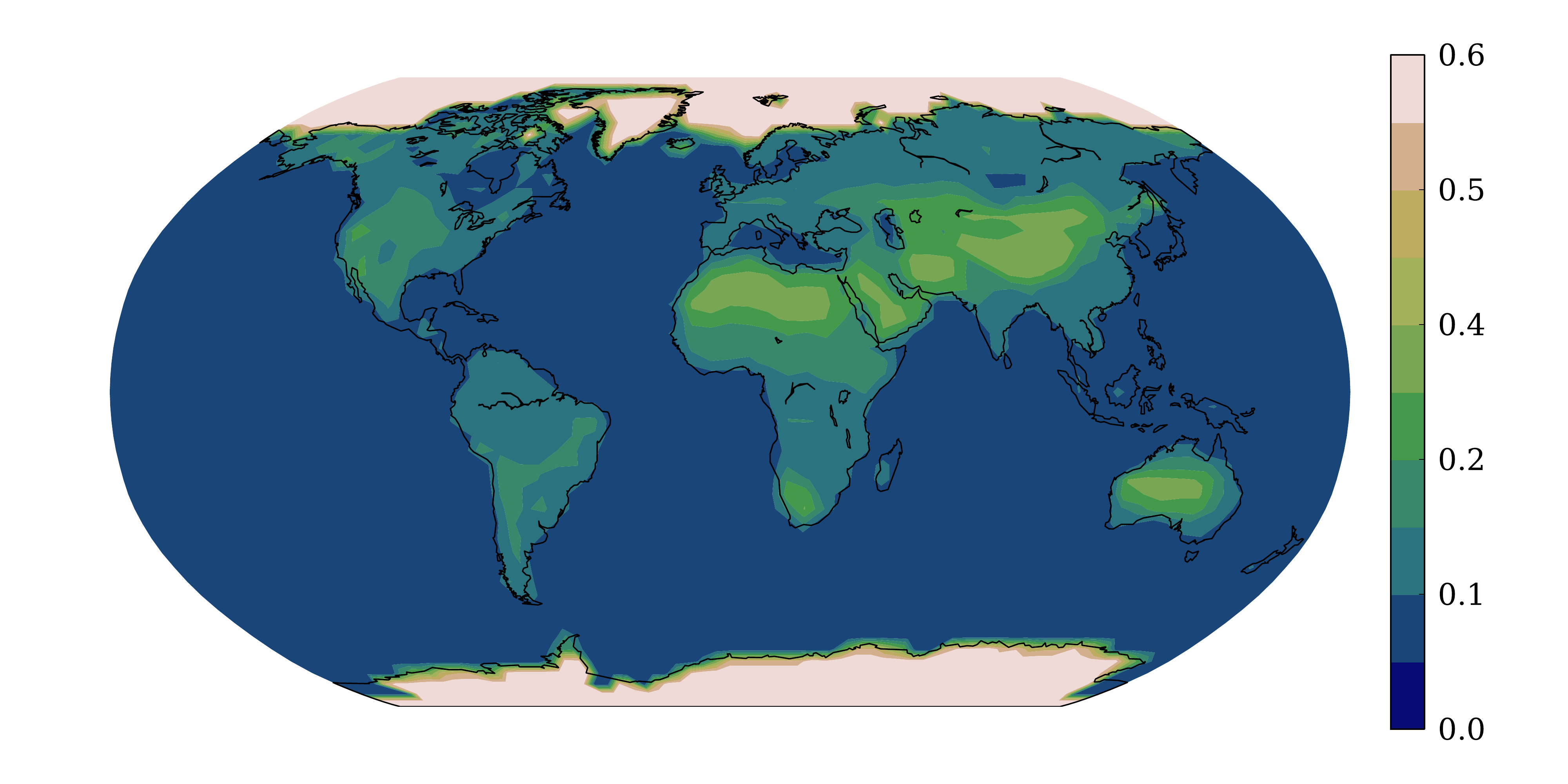}
\noindent\includegraphics[width=20pc]{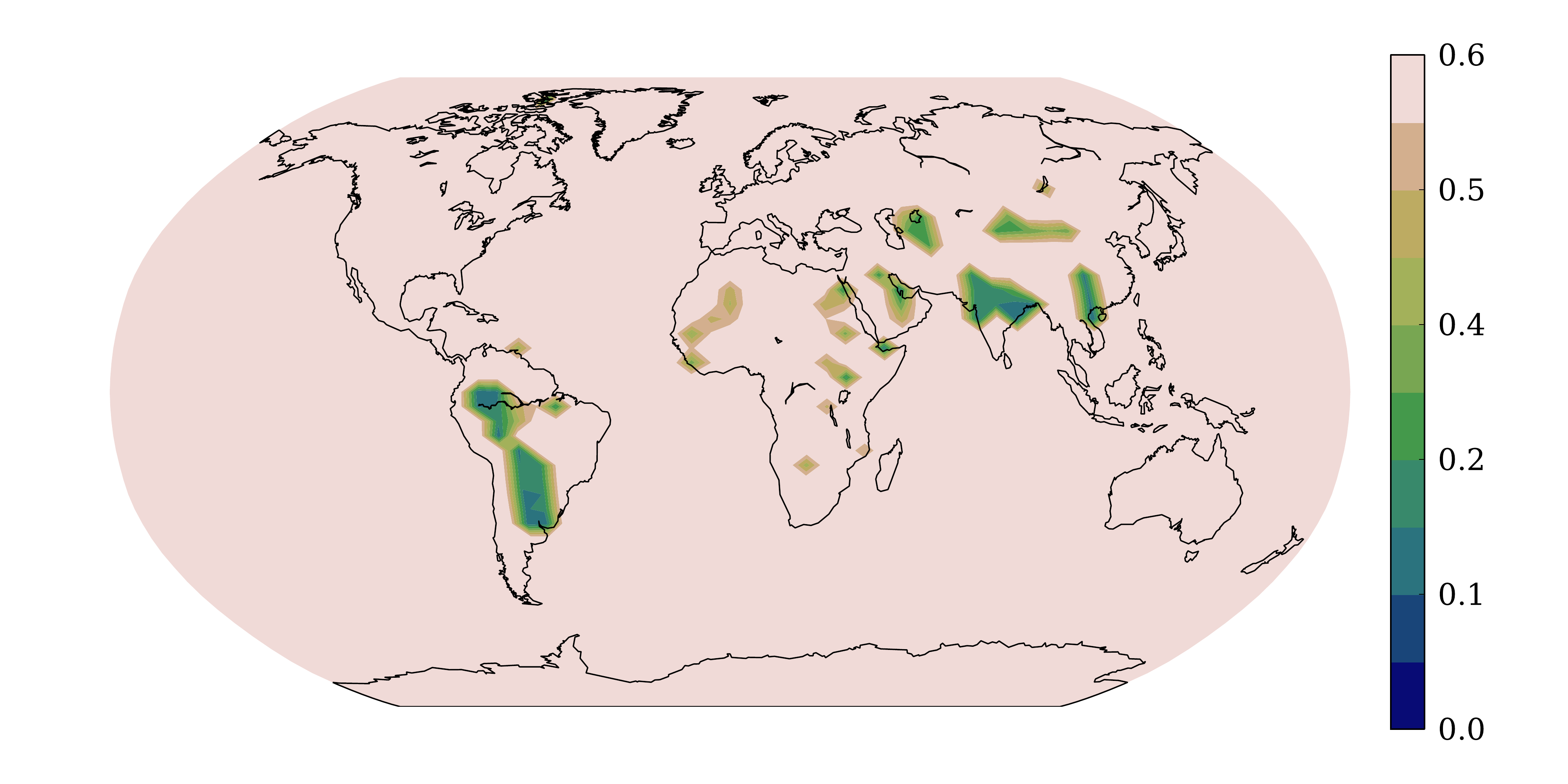}
\noindent\includegraphics[width=20pc]{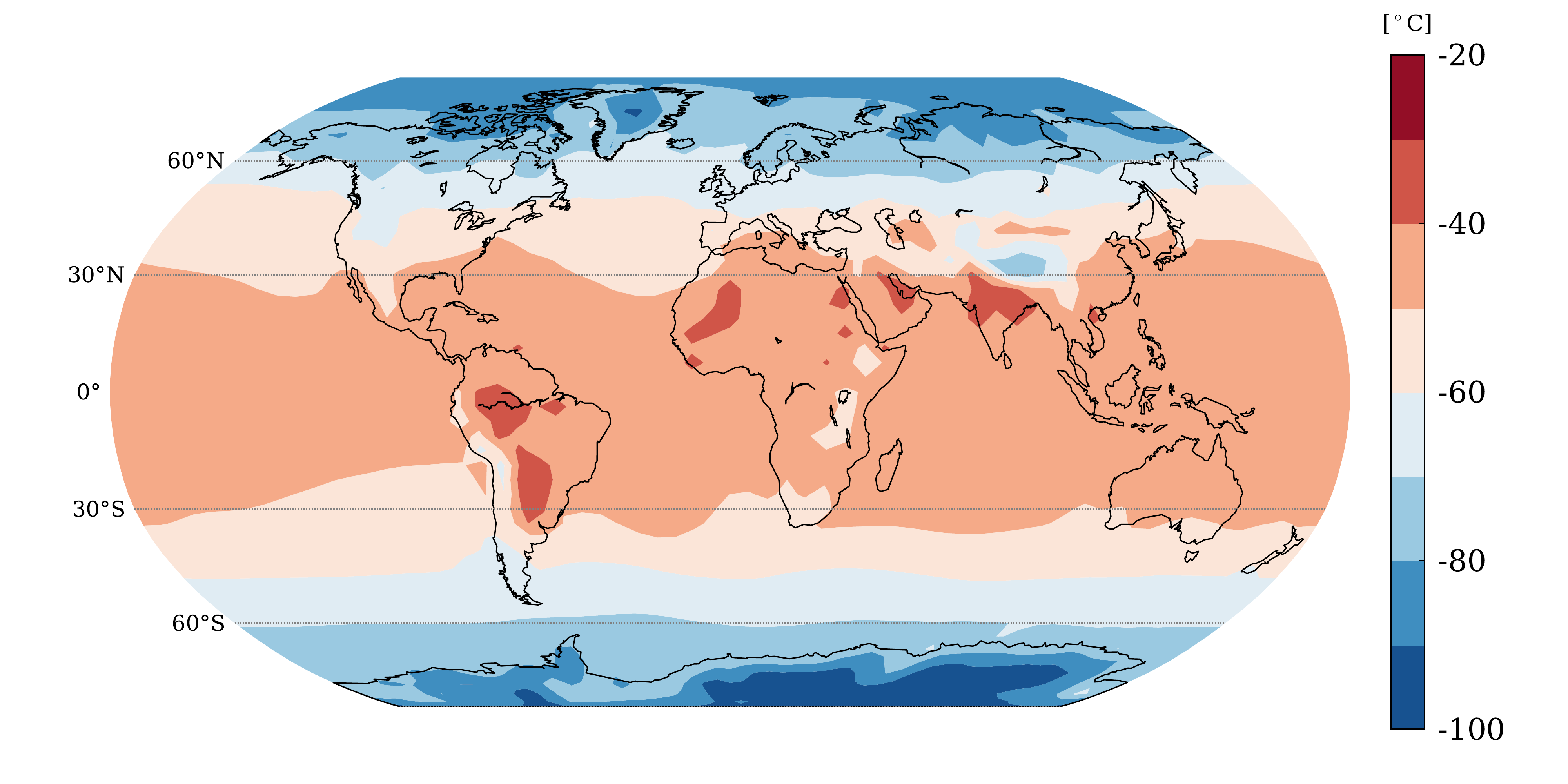}
\caption{Surface albedo at the beginning (top left) and after 23 years (top right) of simulation of the present-day Earth under a 20$\%$ weaker Sun and leading to a full snowball Earth. The bottom panel corresponds to mean surface temperature after 23 years of simulation.}
\label{figure3}
\end{figure*}

First, we ran the model with the present-day continents and atmosphere under a 20$\%$ weaker Sun (corresponding to the Sun 3 Ga ago) to illustrate the faint young Sun problem. 
Figure \ref{figure3} shows the resulting glaciation with the spreading of sea ice from the polar region to the equator. A full "snowball Earth" is produced after only 23 years. The equilibrium mean surface temperature is -54$^\circ$C. Moreover, if we reset the solar insolation to its modern value, the Earth does not exit from the snowball state and remains entirely frozen.

\subsection{Choice of land distribution, greenhouse gases and solar irradiances}
The fraction of the Earth covered by land was different during the Archean than at the present time.
The continental crust volume was lower and has increased during the Archean \cite[]{belousova10, dhuime12}. Thermal modelling and hypsometry indicate that most continents were flooded throughout the Archean \cite[]{flament08}. In addition,  the ocean volume may have been up to 25$\%$ greater than today \cite[]{pope12}. Thus, it is realistic to believe that the fraction of emerged land was low. Lands could have emerged essentially at the end of the Archean \cite[]{flament08}. This has been suggested as the key element triggering the Great Oxidation Event, associated with the change in the oxidation state of volcanic gases transiting from submarine to aerial volcanism \cite[]{kump08b, gaillard12}. In this study, we consider three periods of the Archean for which we run simulations: the beginning at 3.8 Ga, the middle at 3 Ga and the end at 2.5 Ga. For simplicity, we make the hypothesis that land appears after 3 Ga and that there was little land before. Thus we run the model with no land for 3.8 and 3 Ga and with a flat equatorial supercontinent covering 20 $\%$ of the Earth at 2.5 Ga. An equatorial supercontinent is the ideal case to keep ice-free land.  If ice-free land cannot be maintained with a supercontinent at the equator, then it would be the same at any location. The boundaries of the supercontinent are +/- 38$^\circ$ in latitude and +/- 56$^\circ$ in longitude, and the altitude is zero for simplicity. The albedo of Archean land is unknown. We have chosen an albedo of 0.3, which is relatively large even for a surface without vegetation (for comparison, the Moon's albedo is under 0.15). This maximizes the cooling effect of land and gives us robust conclusions concerning the impact of land distribution.
We discuss this in the next section.

Concerning the atmosphere, most of the simulations are performed with one of the following 3 atmospheric compositions:

\begin{itemize}
\item Composition A: a low level of CO$_2$ (0.9 mbar of CO$_2$ and CH$_4$) corresponding to the composition from \cite{rosing10}
\item Composition B: an intermediate level of CO$_2$ (10 mbars of CO$_2$ and 2 mbars of CH$_4$) corresponding approximately to the maximal amount of CO$_2$ and CH$_4$ consistent with the geological constraints at the end of the Archean (see introduction)
\item Composition C: a high level of CO$_2$ (0.1 bar of CO$_2$ and 2 mbars of CH$_4$)
\end{itemize}

Composition A and B could lead to the formation of organic hazes, yet we ignore their impact.
The mean surface pressure is taken equal to 1 bar for simplicity in all cases, so completing the rest of the atmosphere with N$_2$. Hence, we use a partial pressure of N$_2$ higher than today (0.8 bar). This may have been the case, yet the impact of 0.2 additional bar of N$_2$ remains limited (see section 5.2.).

Concerning the astronomical parameters, we used the present-day obliquity but no eccentricity for the orbit. The solar luminosity $L$ of the Archean Sun has been fixed in our simulation by the formula of \cite{gough81}:

\begin{equation}
L(t)=\left[ 1+\frac{2}{5} \left( 1-\frac{t}{t_{\odot}} \right) \right] ^{-1} L_{\odot}
\end{equation}

where $t$ is the time before today, $t_{\odot}$=4.57 Gyr is the age of the solar system and $L_{\odot}$=3.85$\times$10$^{26}$ W is the modern solar luminosity.
If we use a solar constant of 1366 W/m$^{2}$ for the modern Sun, we get a solar constant for the Archean Sun of 1120.8, 1081.9 and 1025.1 W/m$^{2}$ at 2.5, 3 and 3.8 Ga respectively.

\subsection{Effect of lands}
\begin{figure*}[h!]
\begin{center}
\hfill \noindent\includegraphics[width=11.5pc]{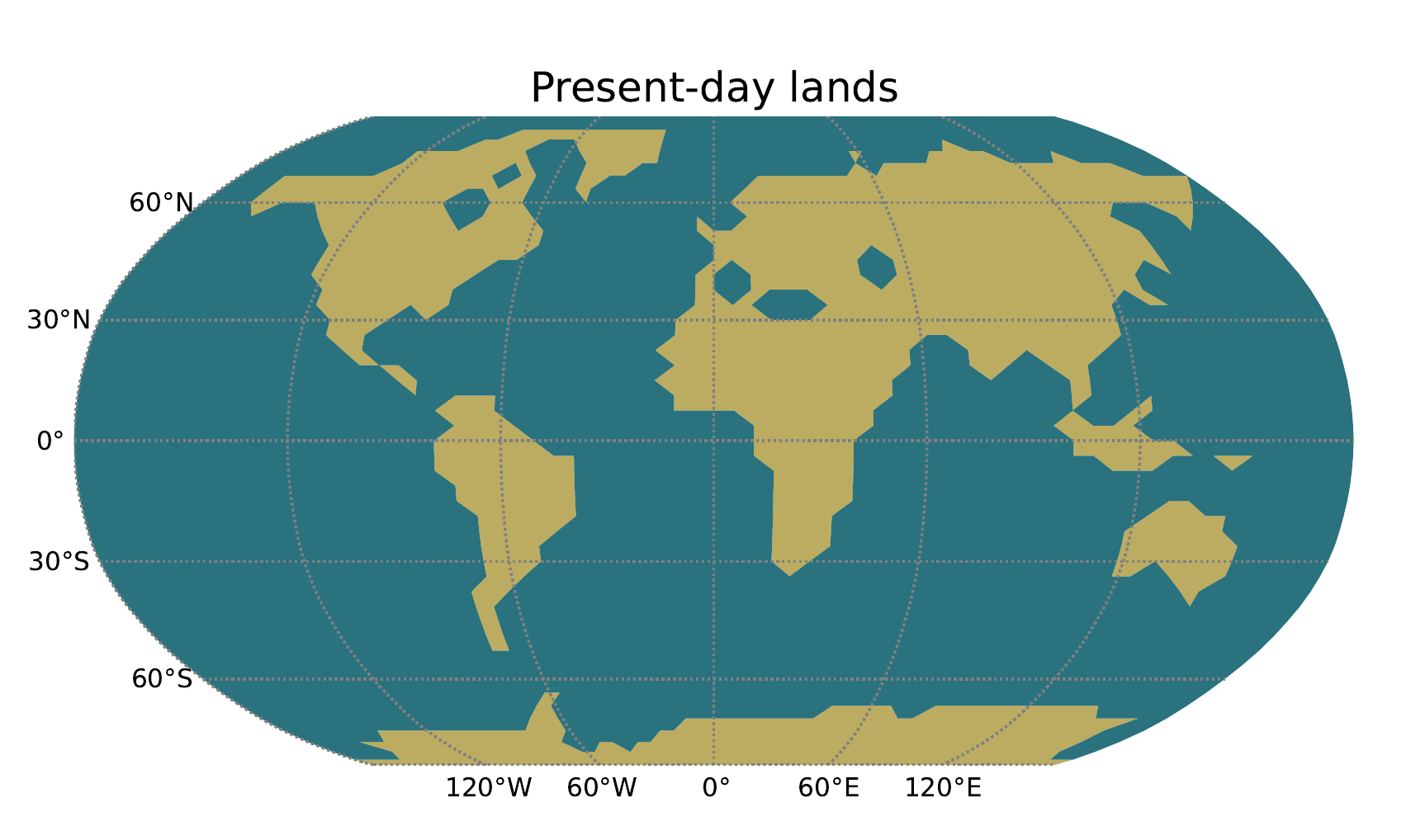}\hfill
\noindent\includegraphics[width=11.5pc]{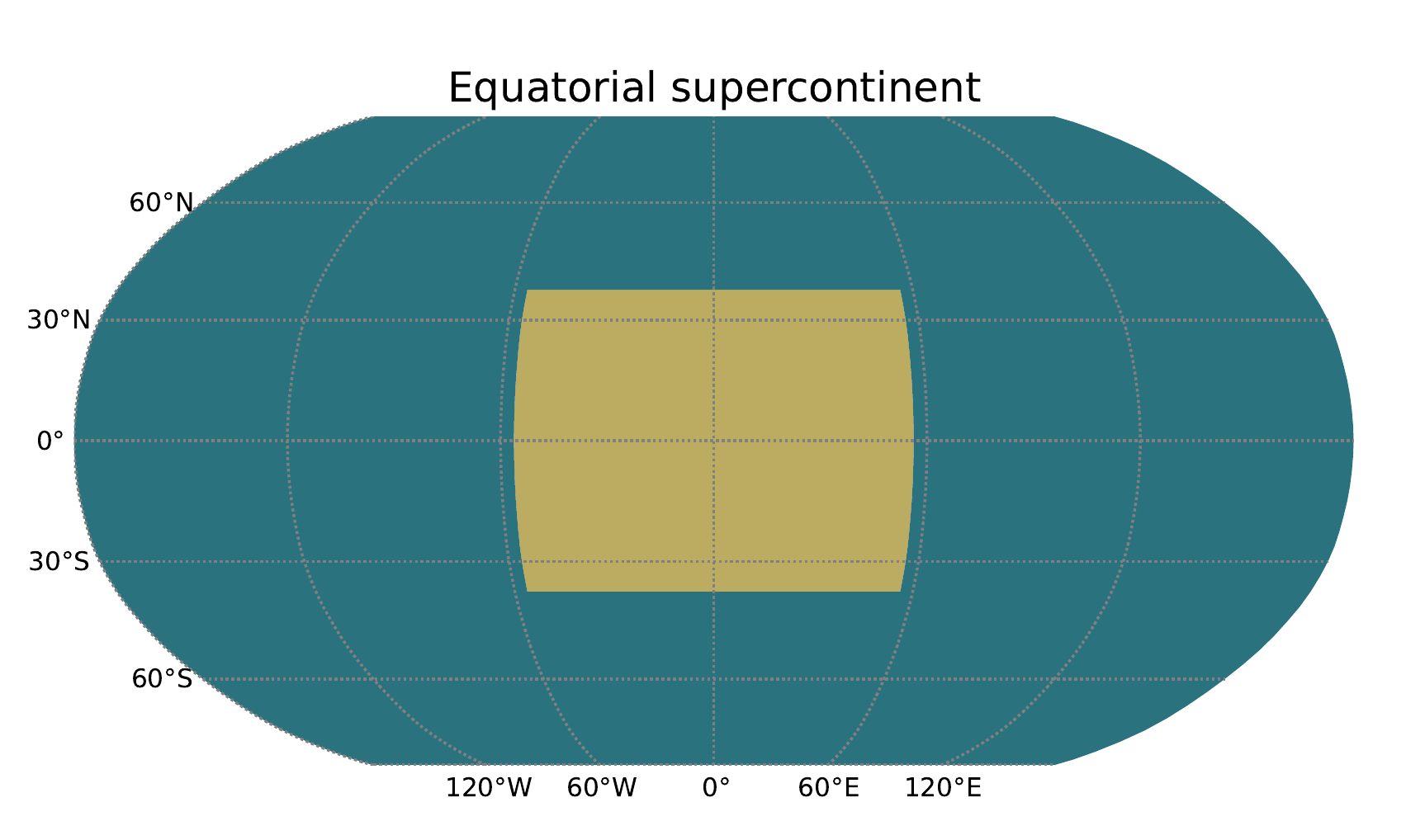}\hfill
\noindent\includegraphics[width=11.5pc]{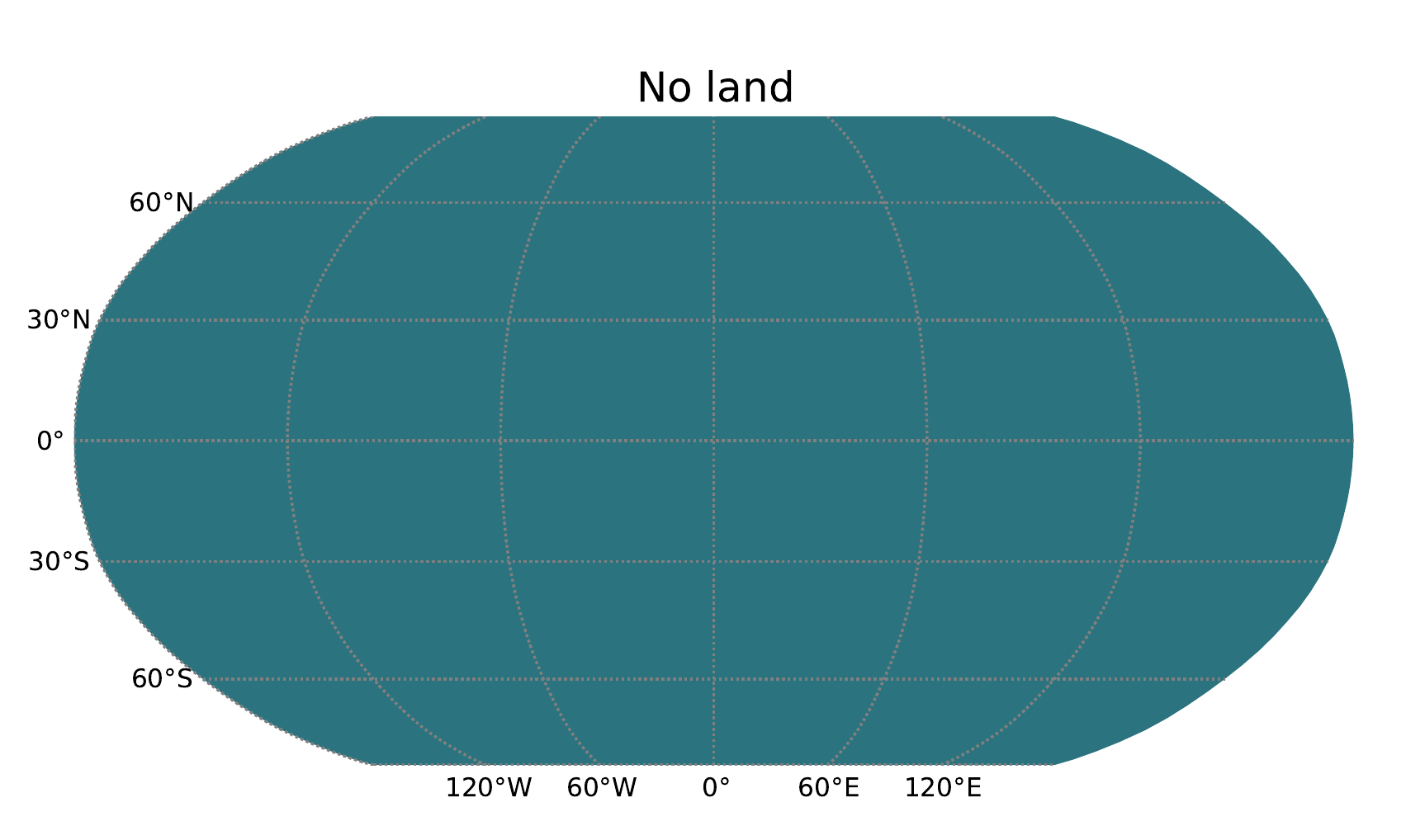}\hfill \hfill \hfill
\end{center}
\begin{center}
\hfill \noindent\includegraphics[width=13pc]{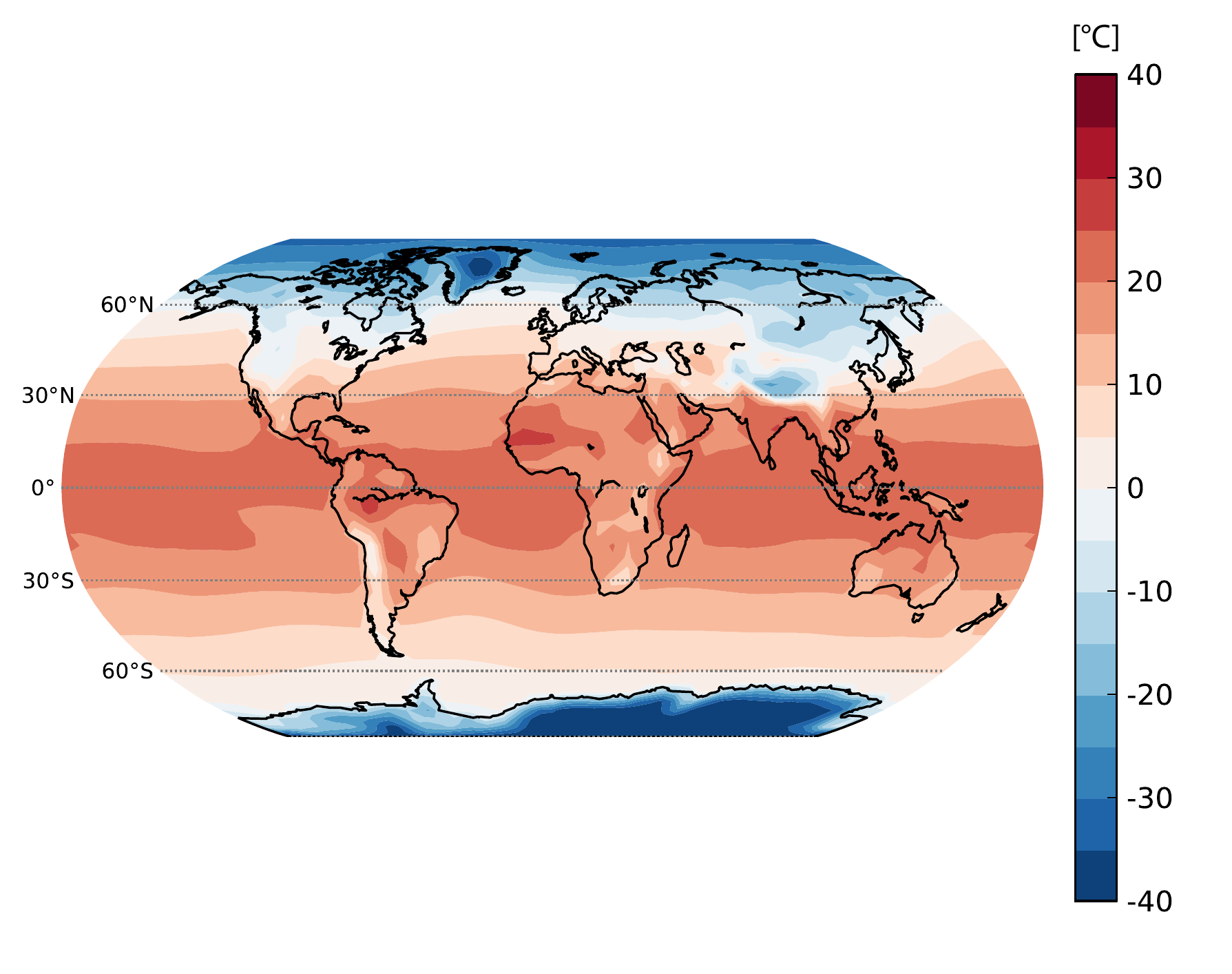}   
\noindent\includegraphics[width=13pc]{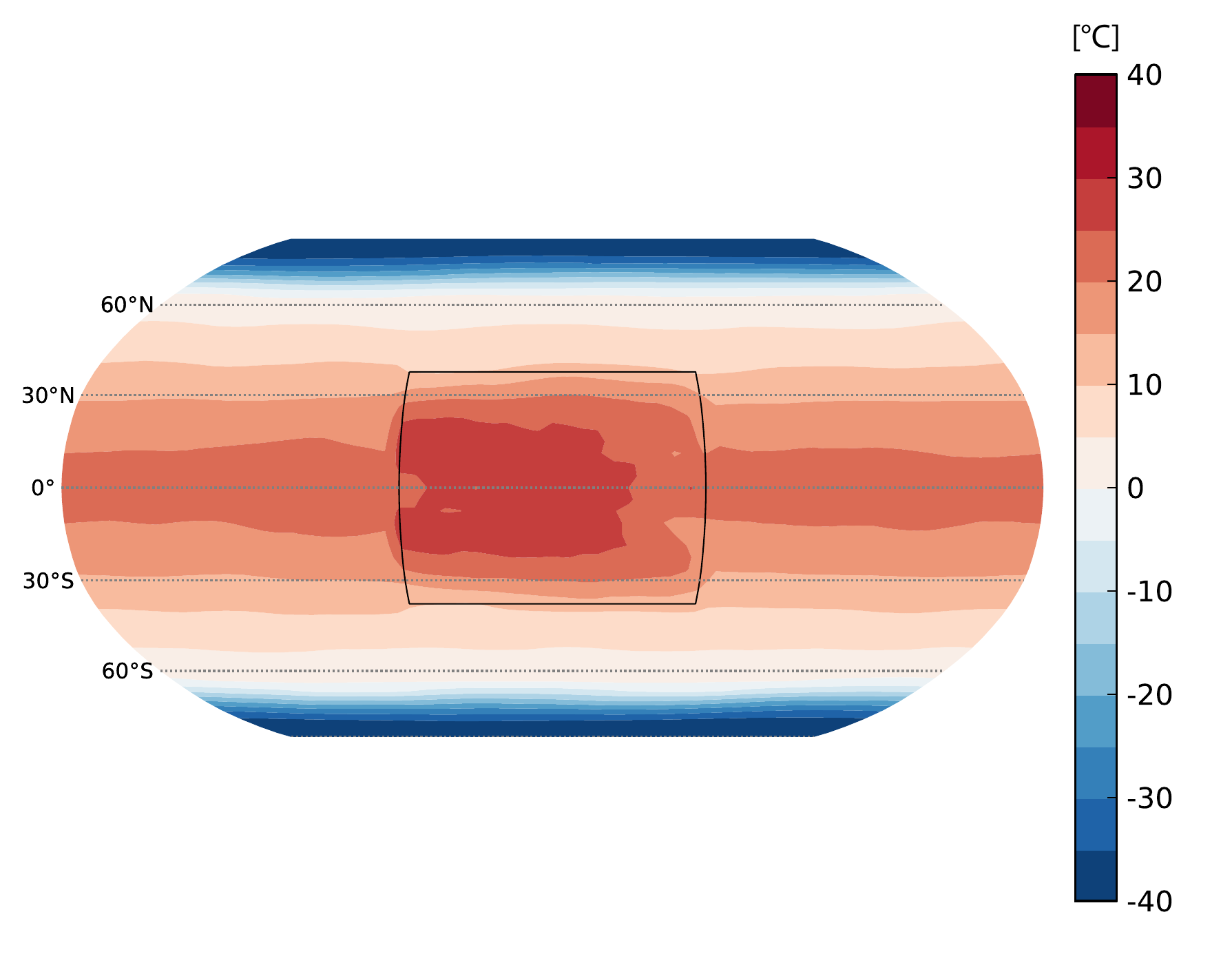}
\noindent\includegraphics[width=13pc]{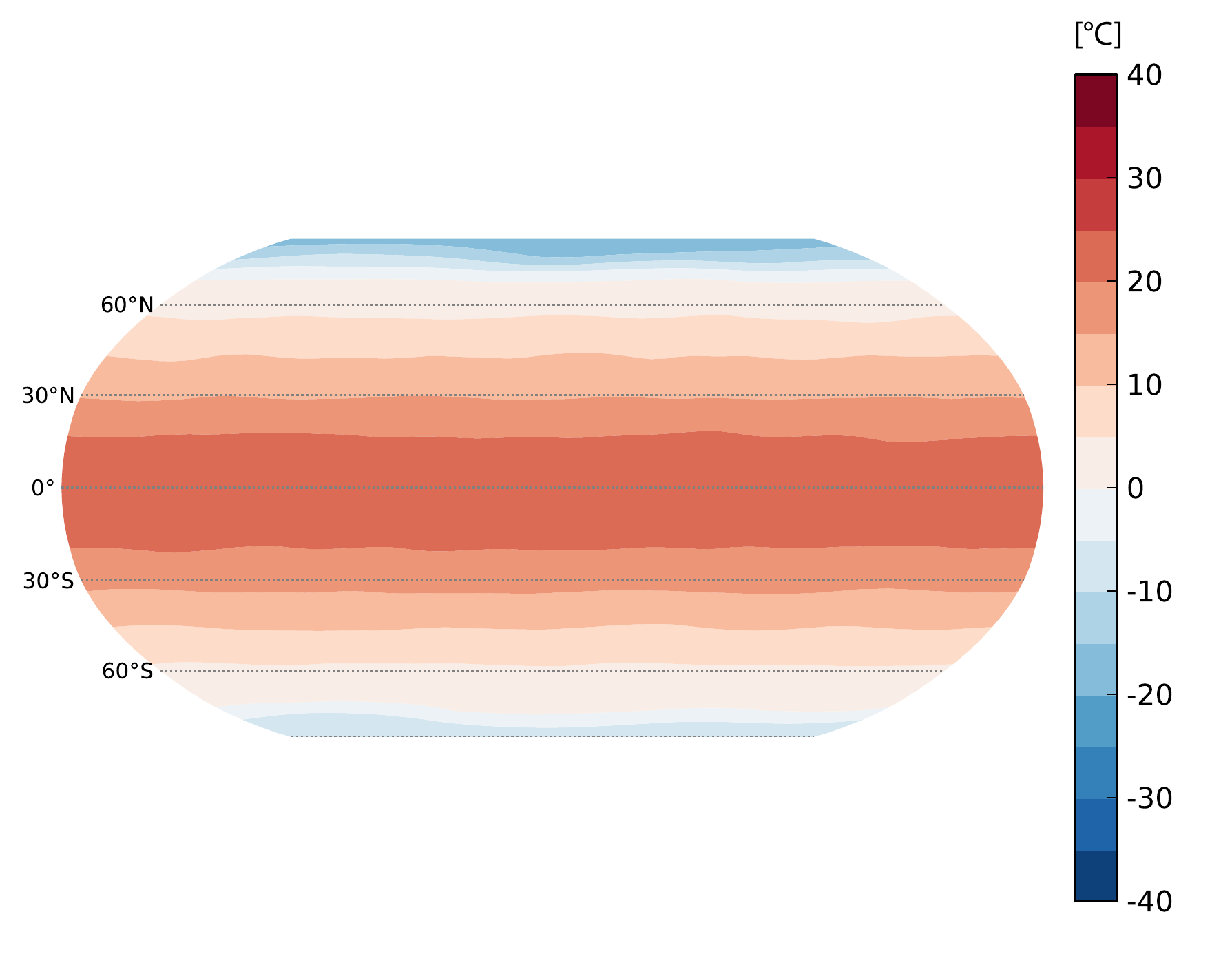}
\end{center}
\caption{Continents (top) and mean surface temperature (bottom) for present-day lands (left), an equatorial supercontinent (middle) and no land at 2.5 Ga (atmospheric composition: 10 mb of CO$_2$ and  2 mb of CH$_4$).}
\label{figure4}
\end{figure*}

According to 1D models, in the absence of organic haze, composition B is sufficient to get a temperate climate at 2.8 Ga \cite[]{haqq-misra08}. Therefore, we chose this composition to assess the influence of emerged land at the end of the Archean at 2.5 Ga. Figure \ref{figure4} shows the annually averaged surface temperatures obtained with either present-day land, the equatorial supercontinent or no land. We get a mean surface temperature of 10.5$^\circ$C with present-day land, 11.6$^\circ$C with the equatorial supercontinent and 13.8$^\circ$C with no land (see table 1). In all cases, we get colder climates than today. For the present-day land, we get more continental ice at high latitudes. Although no traces of glaciation have been observed for most of the Archean, that could be acceptable given the few data available and the impossibility of determining their latitudes. In the simulation with the supercontinent, we get a desert climate with very little precipitation and warm temperatures on the continent. This case would be consistent with the absence of glaciation for most of the Archean \cite[]{kasting06b}. For the Earth with no land, we get a warmer ocean than in the other cases and less sea ice at the poles. We can conclude that land, regardless of its location, has a limited impact on the mean surface temperature. The removal of all land compared to present-day leads a to a global warming of 3.3$^\circ$C (see table 1). The mean tropical oceanic temperature is even less affected (a difference of less than 1.5$^\circ$C).
We found a planetary albedo of 0.34 with present-day land and 0.33 for the supercontinent and no land. The removal of land produces a radiative forcing of $\sim$10 W/m$^2$ for present-day land and only $\sim$2 W/m$^2$ for the supercontinent. Thus the planetary albedo is approximately the same with the equatorial supercontinent than with no land, even assuming a high ground albedo of 0.3.

\begin{figure}[!]
\centering
\noindent\includegraphics[width=20pc]{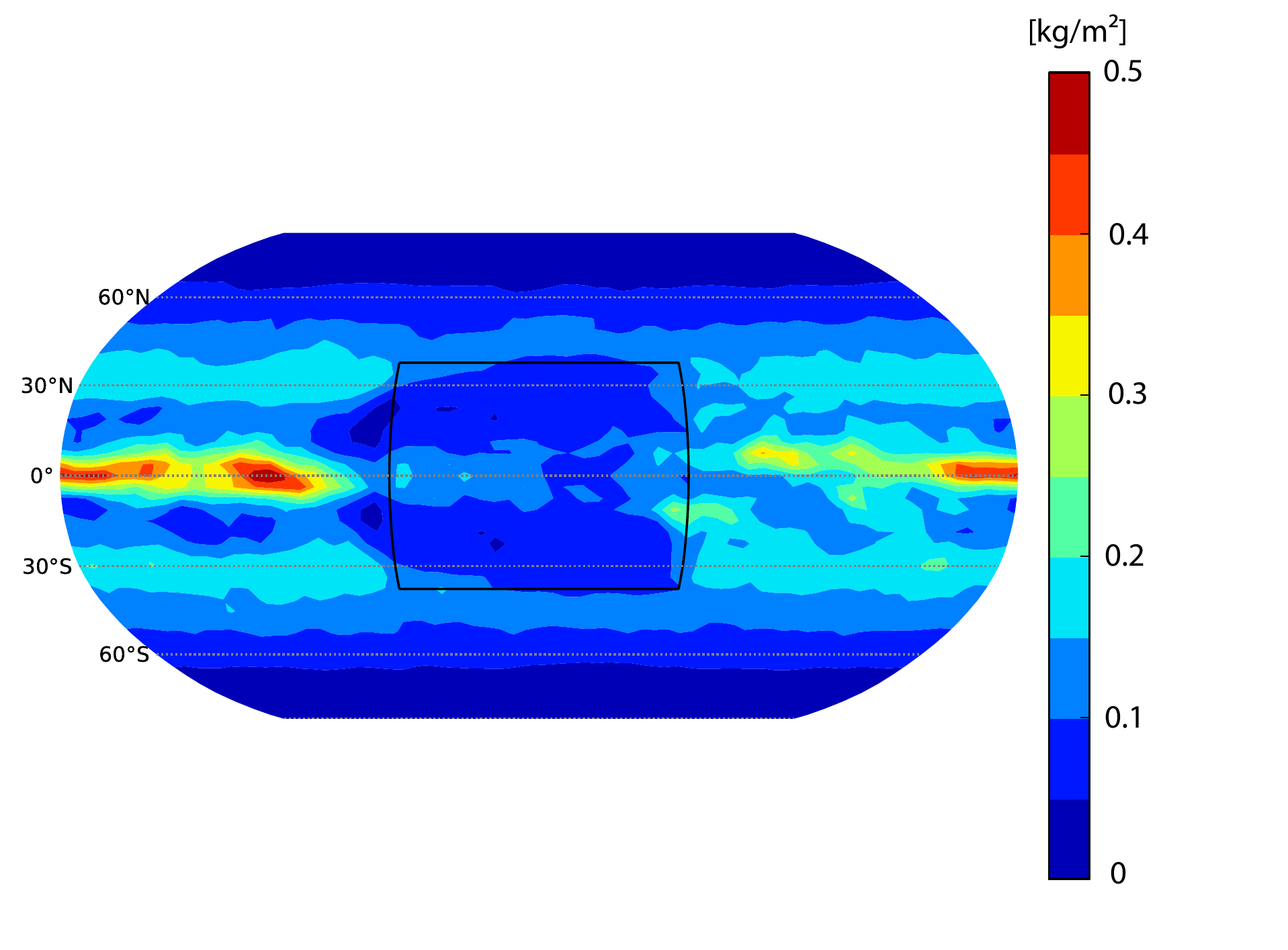}
\caption{Column of condensed water (liquid and icy clouds) with an equatorial supercontinent at 2.5 Ga (atmospheric composition: 10 mb of CO$_2$ and  2 mb of CH$_4$).}
\label{figure5}
\end{figure}

Contrary to what was deduced from 1D models \cite[]{rosing10}, the presence of land, not covered with ice, does not necessary increase the planetary albedo. There are generally less clouds over lands than over oceans. This is observed on the present-day Earth with less clouds in the northern hemisphere than in the southern hemisphere, and this is clear in our simulations in the case of the supercontinent. Figure \ref{figure5} shows the lack of clouds over the supercontinent. This is caused by the lower evaporation over lands and the subsidence zones forming over them. Yet, even with a similar planetary albedo, the absence of land leads to a higher mean surface temperature owing to the larger amount of water vapor in the atmosphere resulting from more evaporation (see table 1). This larger amount of water vapor also enhanced transport of energy to the poles which become warmer (compare surface temperatures for the case with the supercontinent and the case with no land in Figure \ref{figure5}).

\subsection{Effect of greenhouse gases}
\begin{figure}[h!]
\centering
\noindent\includegraphics[width=20pc]{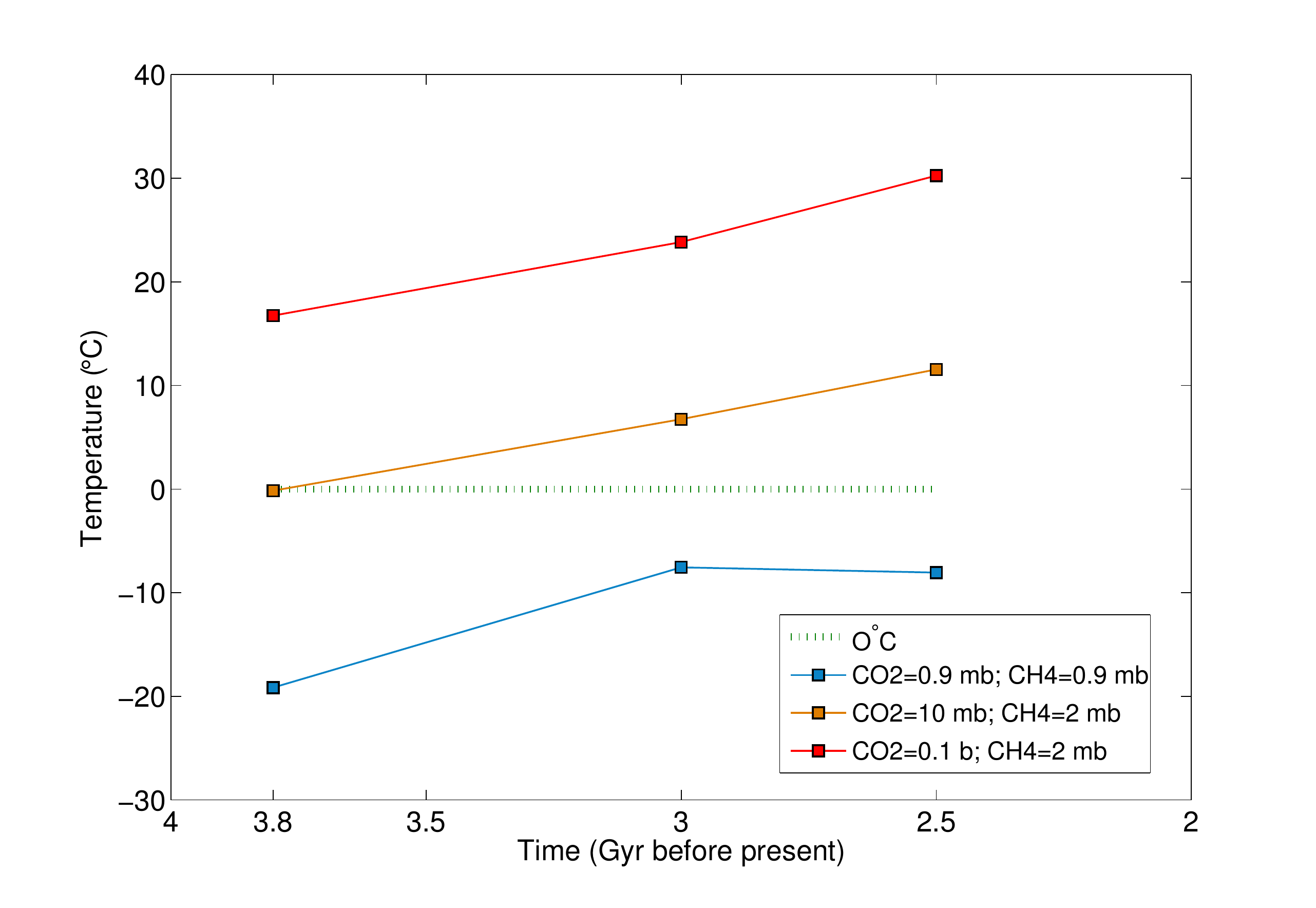}
\caption{Mean surface temperature during the Archean with 0.9 mb of CO$_2$ and CH$_4$ (red line), 10 mb of CO$_2$ and 2 mb of CH$_4$ (orange line), and 0.1 bar of CO$_2$ and 2 mb of CH$_4$ (red line). Dotted green line corresponds to the freezing point of water.}
\label{figure6}
\end{figure}

Figure \ref{figure6} shows the mean surface temperature obtained with our model throughout the Archean with the three atmospheric compositions defined previously. The simulations have been run with the hypothesis for lands from section 4.3., so no land at 3.8 and 3 Ga, and a supercontinent at 2.5 Ga. Composition B, which gives climates close to the present-day for the end of the Archean, is clearly not sufficient to get temperate climates at the beginning of the Archean  (mean surface temperature around 0$^\circ$C at 3.8 Ga). A stronger greenhouse effect is required. Composition C, with 0.1 bar of CO$_2$, allows for a temperate climate, even warmer than today, in the early Archean (mean surface temperature around 17$^\circ$C at 3.8 Ga). Composition A yields mean surface temperature below the freezing point and will be studied in the next section.

\begin{figure}[h!]
\centering
\noindent\includegraphics[width=20pc]{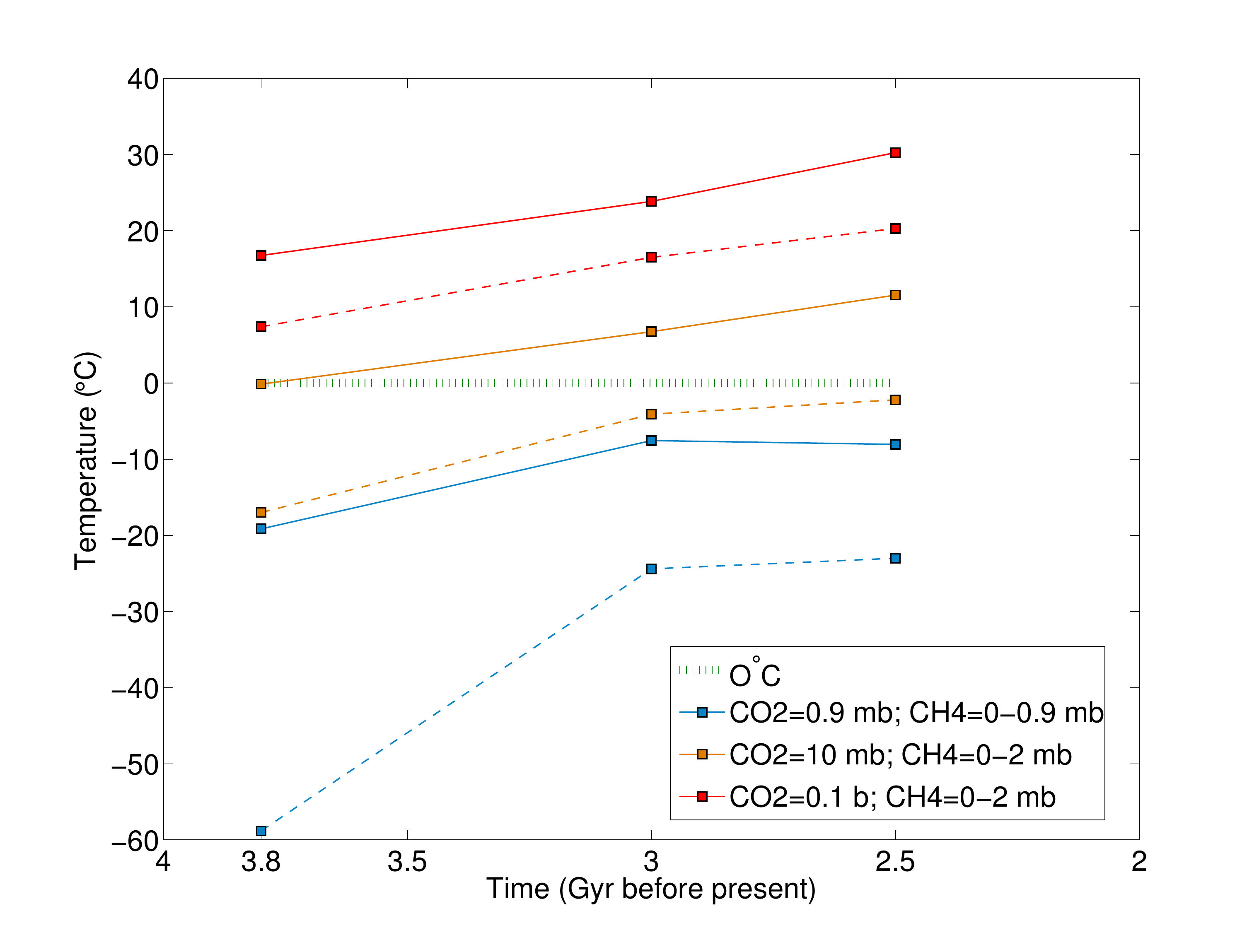}
\caption{Same as figure 6. Dashed lines correspond to atmospheric composition without methane.}
\label{figure7}
\end{figure}

At 2.5 Ga, the lack of methane produces a cooling of 10$^\circ$C for composition C and 14$^\circ$C with a mean surface temperature of  -2.2$^\circ$C for composition B (Figure \ref{figure7}).

These results provide new estimates for the composition of Archean atmosphere. If the climate was temperate (mean surface temperature between 10$^\circ$C and 20$^\circ$C) and if no other warming process was present, a minimum of 0.1 bar (0.01 bar) of CO$_2$ with CH$_4$ at the beginning (the end) of the Archean should have been present in the atmosphere, according to our simulations. Such amounts of CO$_2$ and CH$_4$  remain acceptable given the current geological constraints and are consistent with amounts determined by 1D models \cite[]{haqq-misra08, kasting06a}.
The presence of methane allows a temperate climate in the absence of haze cooling effect. If methane was present in large quantities (around 2 mbars) at the end of the Archean, the oxidation of the atmosphere in the early Proterozoic and the destruction of methane would have produced a strong glaciation with a mean surface temperature possibly below the freezing point, but not necessarily a full glaciation (see next section). This scenario could explain the evidence for glaciation in the early Proterozoic \cite[]{kasting06b}.

\begin{figure}
\centering
\noindent\includegraphics[width=20pc]{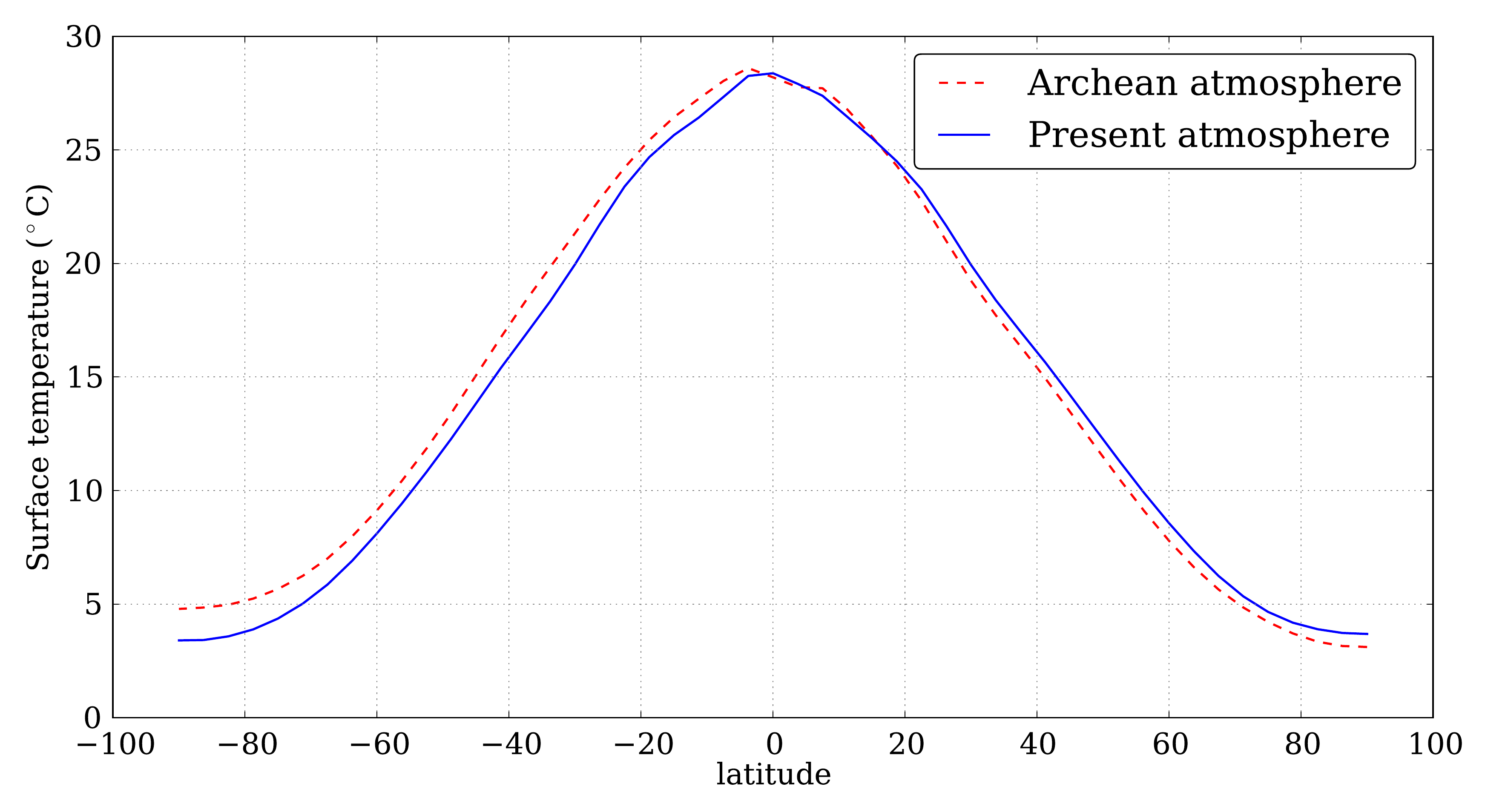}\\
\noindent\includegraphics[width=20pc]{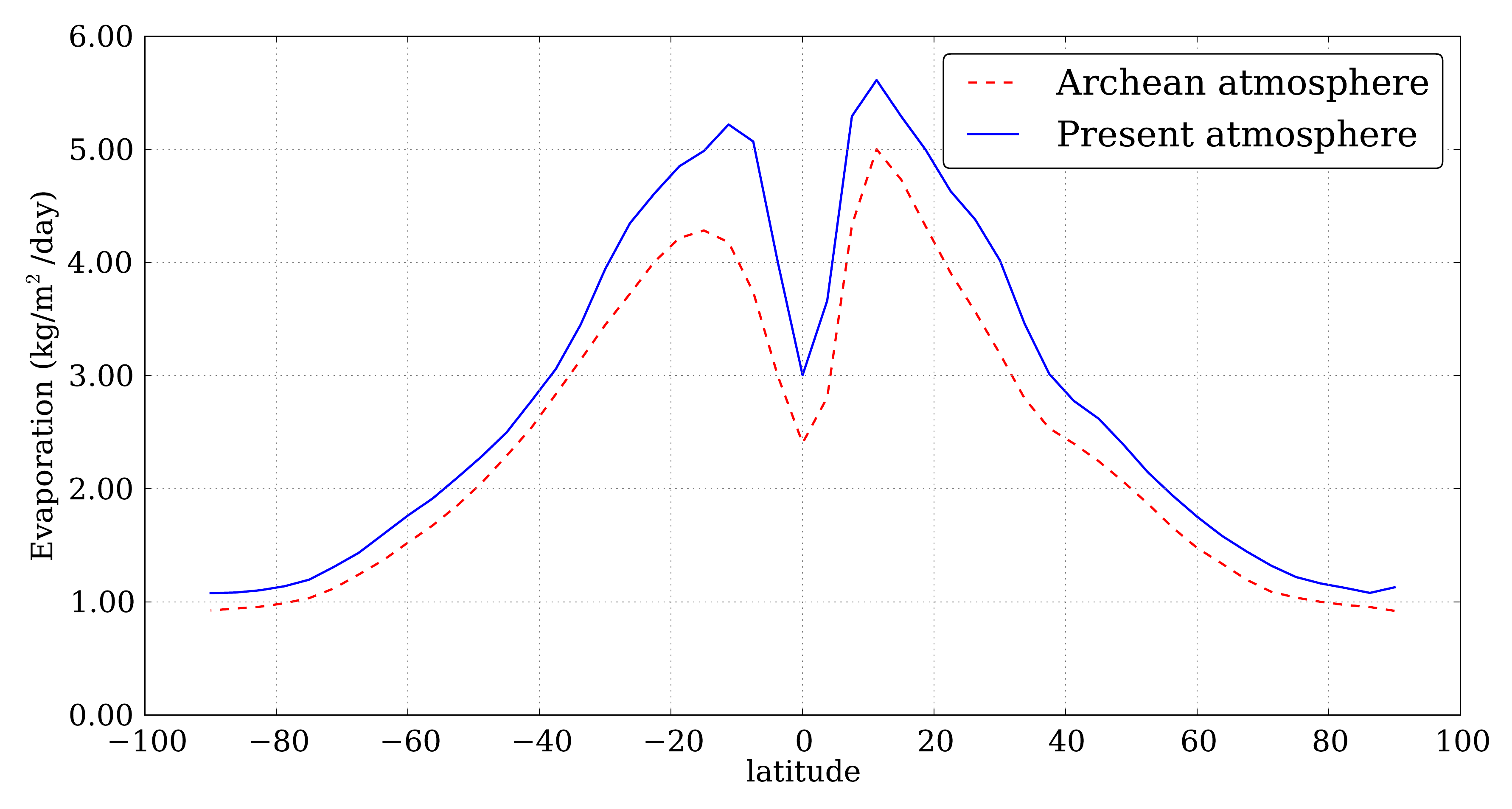}
\caption{Zonally averaged surface temperature (top) and evaporation rate (down) for an aqua-planet with present-day atmospheric composition and Sun (red) and for an aqua-planet with 0.1 bar of CO$_2$, 2 mb of CH$_4$ and the archean Sun at 3.8 Ga (red).}
\label{figure8}
\end{figure}

To better understand the differences between the present-day climate and the archean climates, we need to compare simple cases (with no land) with the same mean surface temperature. Thus, we ran the model for an aqua-planet with the modern atmospheric composition and solar irradiance. This case has approximately the same mean surface temperature as the Earth at 3.8 Ga (still no land) with the atmospheric composition C (see table 2 and Figure \ref{figure8}). The comparison between the two emphasizes the key characteristics of archean climates, with reduced Sun and enhanced greenhouse effect.

Our simulations predict a strong decrease in the planetary albedo for the Archean Earth (passing from 0.36 to 0.3 in table 2). This occurs for two reasons.
First, the enhancement of greenhouse gases, in particular methane, increases the shortwave solar radiation absorbed by the atmosphere. $\sim$25$\%$ of the incoming solar flux is absorbed by the atmosphere for the Earth at 3.8 Ga with composition C versus 19$\%$ for modern Earth. 
It contributes to approximately half the decrease in the planetary albedo and warms the planet.

Second, the lower solar radiation absorbed by the oceans leads to less clouds. The hydrological cycle is weakened with less evaporation across the planet (see Figure \ref{figure8}). Less clouds are produced (both lower and higher clouds), thereby reducing the planetary albedo. The column mass of clouds is reduced by $\sim$16$\%$ and the cloud cover by $\sim$10$\%$. This tends to globally warm the planet. Precipitation is also reduced by $\sim$15$\%$, which would diminish a little the weathering for the Archean Earth.

These differences (warmer poles, a reduced planetary albedo, less clouds and precipitation) are very general and have to appear in any temperate climate of the early Earth.

\subsection{Case of a cold climate}
\begin{figure}
\centering
\noindent\includegraphics[width=20pc]{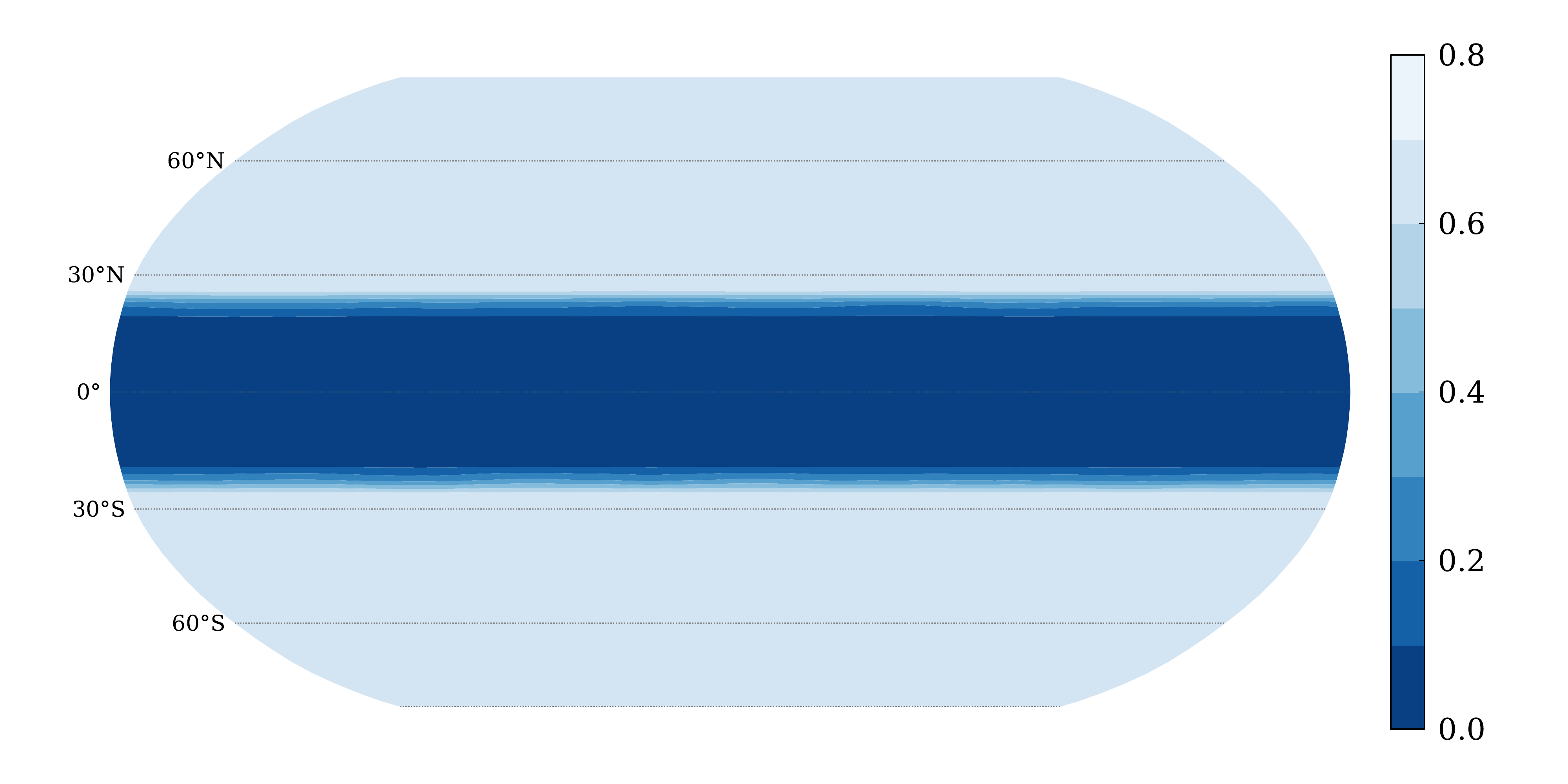}
\caption{Surface albedo at 3 Ga with no land (atmospheric composition: 0.9 mb of CO$_2$ and no CH$_4$).}
\label{figure9}
\end{figure}

Here, we provide more details on the Archean climates for which the mean surface temperature is below the freezing point. That corresponds to very cold climates with the atmospheric composition B without methane, and composition A with or without methane (Figure \ref{figure7}). According to 1D models, all these cases should correspond to a full glaciation, while in our simulations, only composition A without methane at 3.8 Ga produces a full snowball Earth. The mean surface temperature can be as low as -25$^\circ$C keeping an ice-free waterbelt at the equator (see figure \ref{figure9}). The sea ice can spread down to 25$^\circ$ N/S without triggering the runaway glaciation. In the coldest case that is not fully frozen (0.9 mbar of CO$_2$ at 3 Ga), the polar temperatures vary between -45$^\circ$C and -95$^\circ$C and the equatorial oceanic temperature is around 6$^\circ$C. 

\begin{figure}
\centering
\noindent\includegraphics[width=20pc]{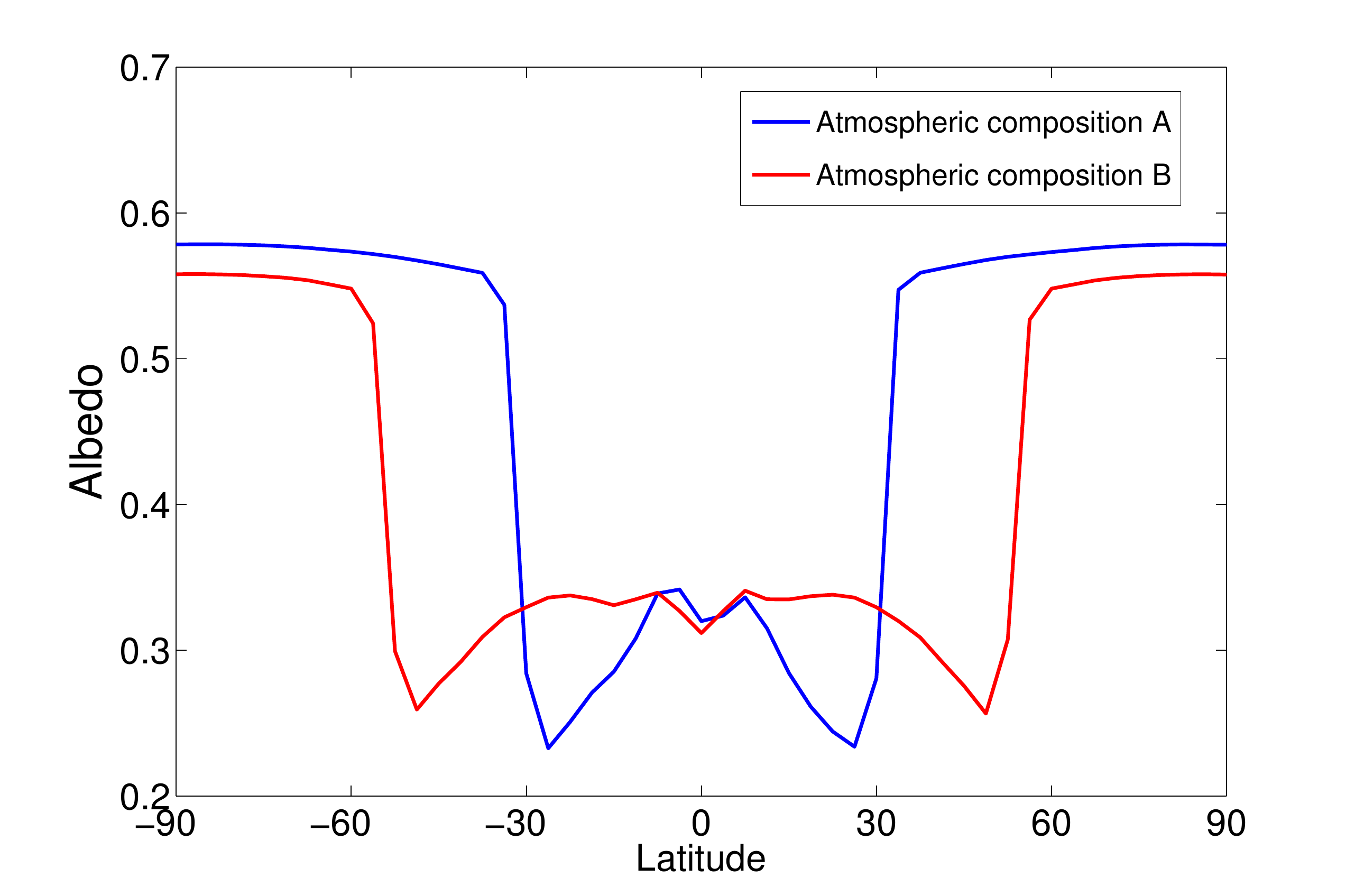}
\caption{Top of the atmosphere albedo at 3.8 Ga with no land with the atmospheric composition A (blue) and B (red).}
\label{figure10}
\end{figure}

In our simulations, we notice a decrease of the albedo above the waterbelt.
In a cold climate, the humidity decreases, leading to thinner clouds and hence a lower albedo. 
This process appears particularly efficient close to the frozen line. The latter corresponds to a cloud-free zone of subsidence. The closer the frozen line goes to the equator, the more the albedo of the tropics decreases. Since this region receives 50$\%$ of Earth's solar insolation, a powerful cloud-albedo feedback counteracts  the ice-albedo feedback and stops the glaciation. Figure \ref{figure10} shows the albedo for two cold climates at 3.8 Ga, the first with atmospheric composition A (mean surface temperature=-18$^\circ$C) and the second with atmospheric composition B (mean surface temperature=0$^\circ$C). In the coldest case, the albedo is reduced by 31$\%$ corresponding to 30 additional W/m$^2$ absorbed by the ocean at 25$^\circ$ N/S,  close to the freezing line.

Thus our model shows that, while a full snowball Earth is reached with present-day atmospheric composition (see subsection 4.1.), the Earth appears very resistant to full glaciation if it has a sufficient amount of greenhouse gases (0.9 mbar of CO$_2$ and CH$_4$ or even without CH$_4$ after 3 Ga) thanks to the cloud feedback. 
Simulations with a coupled climate/ice-sheet model, considering a cloud feedback similar to ours, have also shown that equatorial waterbelts can be maintained under cold climates providing a solution to the Neo-Proterozoic glaciations and keeping a refugium for multicellular animals \cite[]{hyde00}. 

Our GCM is rather simplified, especially as far as the ocean is concerned. Our oceanic model includes diffusion and Ekman transport for heat but no transport of oceanic ice. \cite{lewis03} showed that winds, through Ekman transport, could advect sea ice close to the equator helping to trigger the full glaciation . \cite{kienert13} gets the exact opposite result to ours. They found a strong sensitivity of full glaciation to the ice-albedo feedback. In their model, 0.4 bar of CO$_2$ is required to avoid a full glaciation. This is far larger than with 1D models or than with our model where 0.9 mbar of CO$_2$ are sufficient to avoid full glaciation and 0.1 bar of CO$_2$ yields a warmer climate than today. However, while \cite{kienert13} use a full oceanic general circulation model, they use a highly simplified atmospheric model with parameterizations of meridional transport. This approach used by these authors does not include cloud feedback effects on their simulated climate.

Thus, studies with other GCMs and in particular GCMs coupled to full oceanic models will be required to estimate more accurately the limit of runaway glaciation for the Archean Earth. But it is still probable that the Earth could have kept waterbelts with a mean surface temperature below 0$^\circ$C and with a low amount of greenhouse gases, which mitigates the faint young Sun problem.

\section{Additional warming processes}
In the previous section, we explored conditions for a temperate climate, by changing only the land distribution and the atmospheric composition. However, it appears difficult to reach or exceed the present-day temperatures with the geological constraints on greenhouse gas partial pressure. Other mechanisms such as the microphysics of clouds, the atmospheric pressure, or Earth's rotation rate have been suggested to warm the early Earth. In this section, we study these additional warming processes with our 3D model.

\subsection{Effect of the cloud droplet size}
Clouds have a strong impact on the terrestrial radiative budget and a change in their cover or thickness could have had an important effect for the climate of the early Earth \cite[]{goldblatt11}. We will focus on the lower clouds, which have a strong impact on the quantity of shortwave radiation arriving on the surface. The removal of all lower clouds compared to the present-day cover would lead to a radiative forcing of 25W/m$^2$ \cite[]{goldblatt11}, i.e. about half of the deficit of radiative forcing due to the weaker Sun during the late Archean. The quantity of cloud condensation nuclei (CCN) is an important factor for cloud microphysics and can control the size of cloud droplets.  For the same mass of condensed water, larger amounts of CCN leads to smaller droplets.
On the modern Earth an important fraction of the total amount of CCN comes from biomass burning and human-generated pollution. These anthropic aerosols are more concentrated over land, leading to a smaller cloud droplet radius over land (6-10 $\mu$m) than over ocean (10-14 $\mu$m) \cite[]{breon02}.

On the prehuman Earth, the quantity of CCNs should have been controlled by biological activity \cite[]{andreae07}, particularly from the dimethylsulfide by algaes and biological particles (including pollen, microbes, and plant debris). Variations of those biological emissions of CCN could have had an impact on the climate, because a reduction of the amount of CCN lead to thinner clouds and so to a lower planetary albedo. That has been suggested to be an explanation of the warm climates of Cretaceous \cite[]{kump08a}. During the Archean, those biological emissions should have been strongly reduced, which should have led to an important warming \cite[]{rosing10}. \cite{rosing10} suggested that the cloud droplet radius could have reached 17 $\mu$m or larger, leading, for their 1D model, to a radiative forcing of 22 W/m$^2$ and a warming of about +10$^\circ$C. The possibility of reaching such large radii has been debated \cite[]{goldblatt10,goldblatt11} along with the radiative forcing, because of the large uncertainty on the evolution of the precipitation rate with droplet radius.

The visible optical depth of water clouds is given by \cite[]{sanchez-lavega_book}:

 \begin{equation}
 \tau=\frac{3}{2}\frac{w}{ \rho r_e}
 \label{eqlib}
 \end{equation}
where $w$ is the column of liquid water (mass per unit area within the cloud), $\rho$ is the volumic mass of water and $r_e$ is the effective radius of the cloud droplets. Thus larger cloud droplets lead to a lower optical depth in two ways. First, by directly increasing $r_e$ (Twomey effect) \cite[]{twomey77}, and second, by decreasing $w$ due to enhanced precipitation at larger radius (Albrecht effect) \cite[]{albrecht89}.
The first way is very robust while the second is very unclear. Larger droplets should naturally increase precipitation, due to larger cross sections of collisions between droplets yielding a larger autoconversion into raindrops \cite[]{boucher95}. However, the magnitude of this increase in precipitation is not well known. Moreover, unclear feedbacks could occur, limiting this effect. For instance, precipitation changes the number of CCN locally and hence the radius of droplets. Moreover, a change in clouds impacts surface wind speeds and thus the emission of CCN.  Such subtle feedbacks are not taken into account in GCMs, which use fixed radii or fixed amounts of CCN, and simplified parameterizations for rain. This remains an open question.

Parameterizations of enhanced rain-out vary from proportional to $r/r_0$ to proportional to $(r/r_0)^{5.37}$ \cite[]{kump08a,penner06,goldblatt10}. \cite{kump08a} and \cite{rosing10} used an exponent 3, an intermediate value. In our model, we use an exponent 1. This value is in the lower range compared to the literature. However, we believe using a physically-based parameterization such as \cite{boucher95} (see equation (\ref{precip})) is important to get robust results. That is why we take it as a reference. A comparison using our model with different exponents is given later. 

\begin{figure}[h!]
\centering
\noindent\includegraphics[width=20pc]{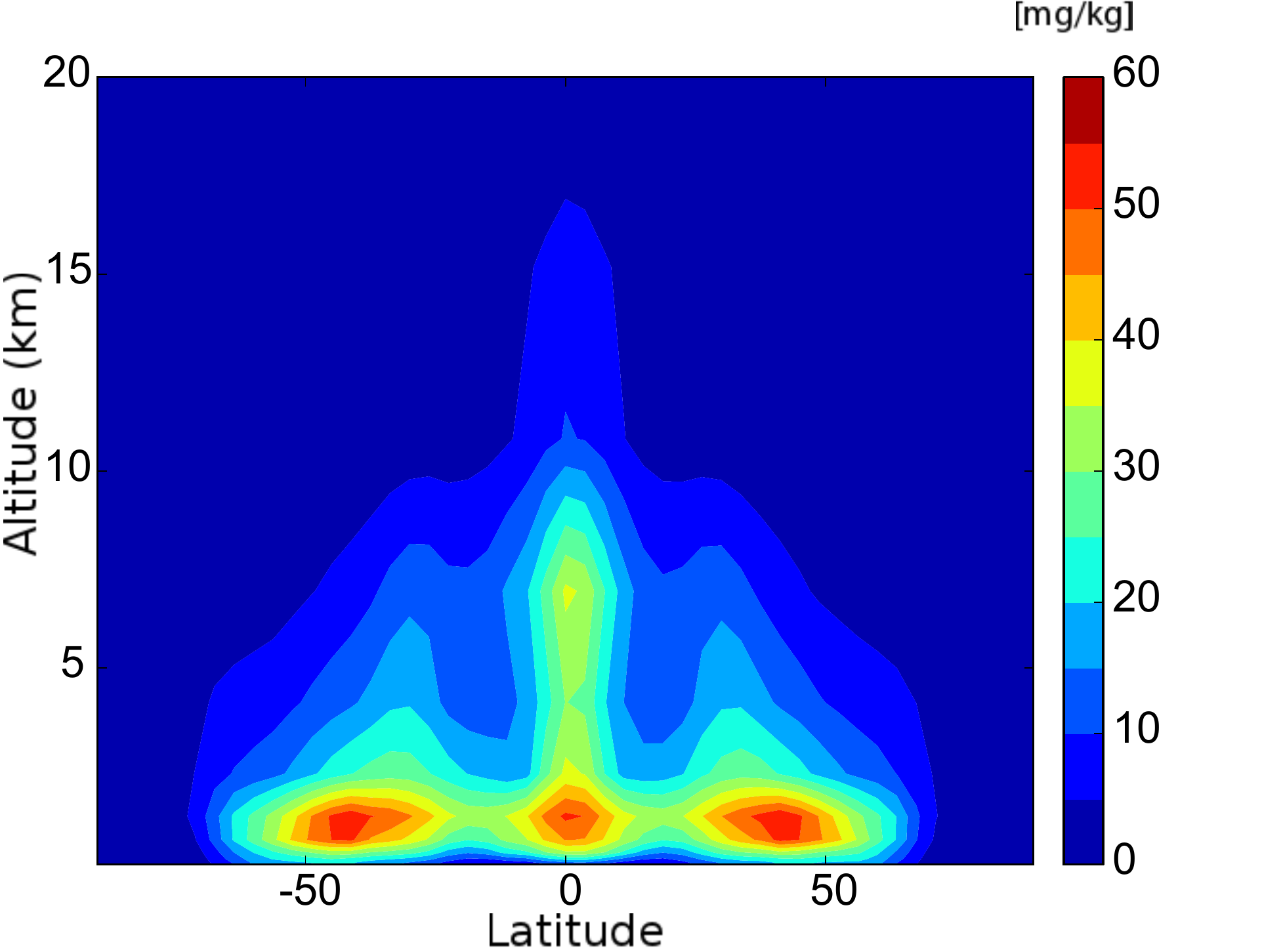}\\
\noindent\includegraphics[width=20pc]{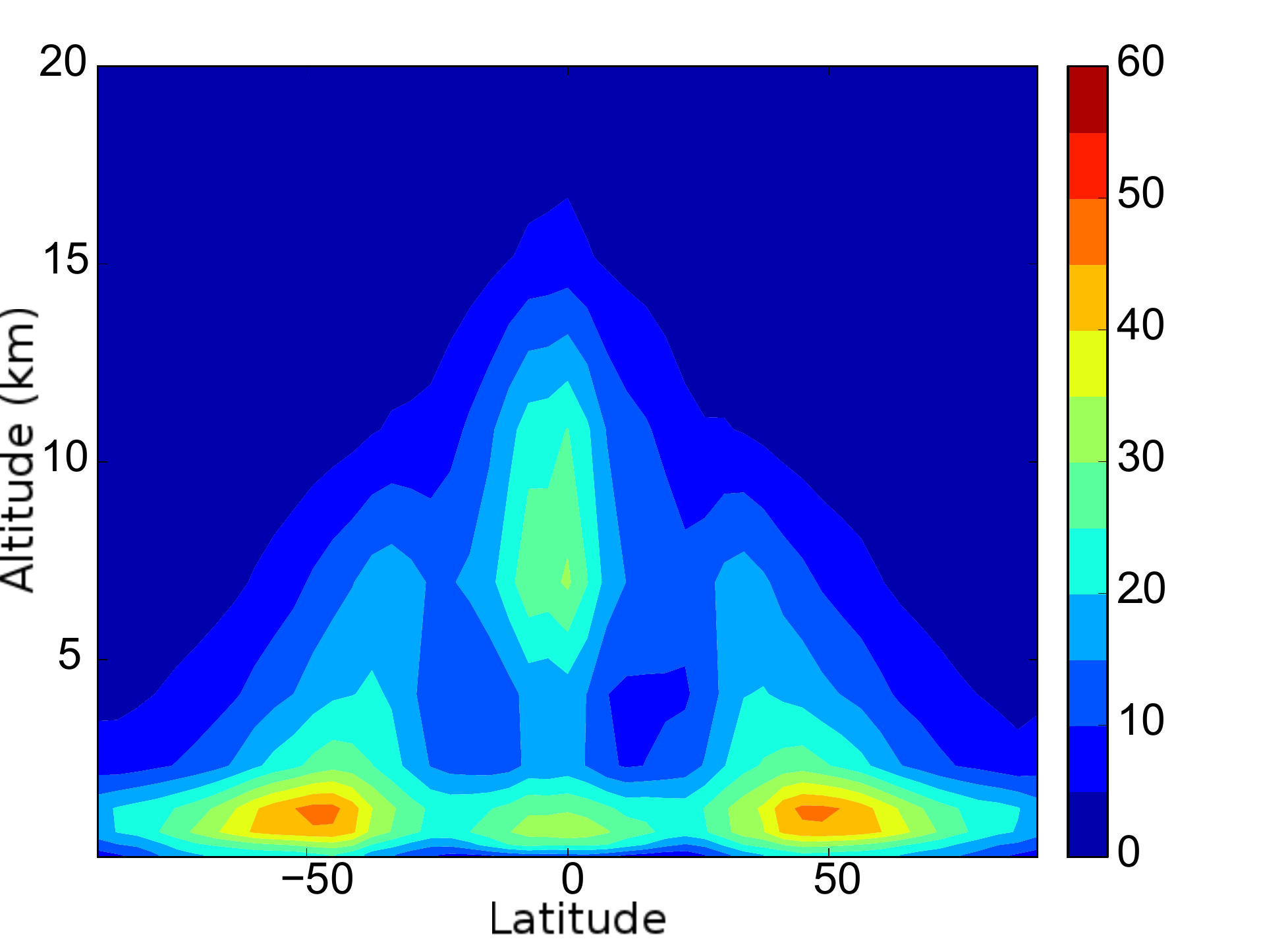}\\
\noindent\includegraphics[width=20pc]{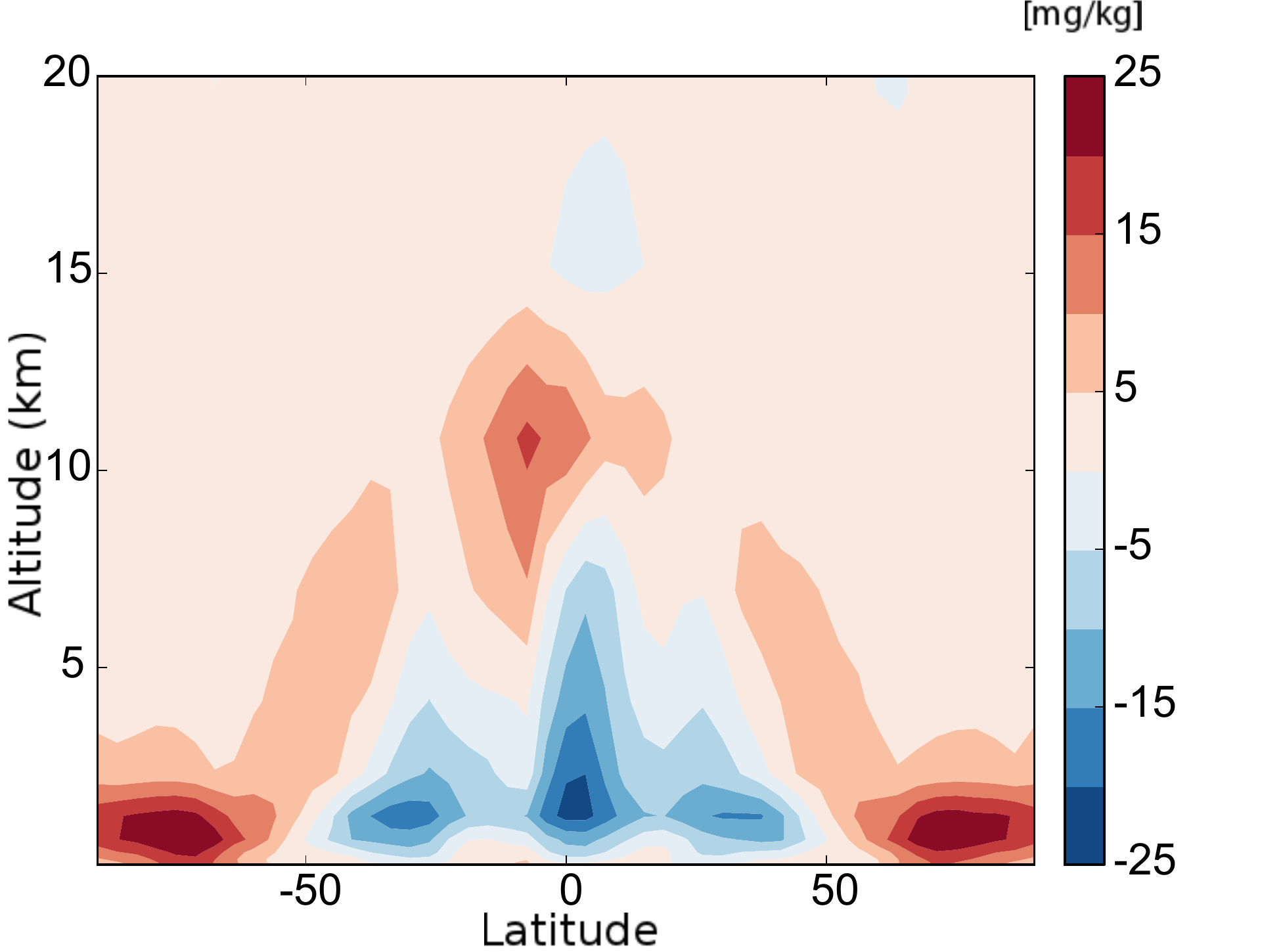}
\caption{Zonally averaged mixing ratio (in mg/kg of air) of condensed water (liquid and icy clouds) at 2.5 Ga (atmospheric composition: 10 mb of CO$_2$ and  2 mb of CH$_4$) for liquid droplet radius of 12 $\mu$m (top) and 17 $\mu$m (middle). The bottom panel is the difference between both (17 $\mu$m minus 12 $\mu$m).}
\label{figure11}
\end{figure}

\begin{figure}[h!]
\centering
\noindent\includegraphics[width=20pc]{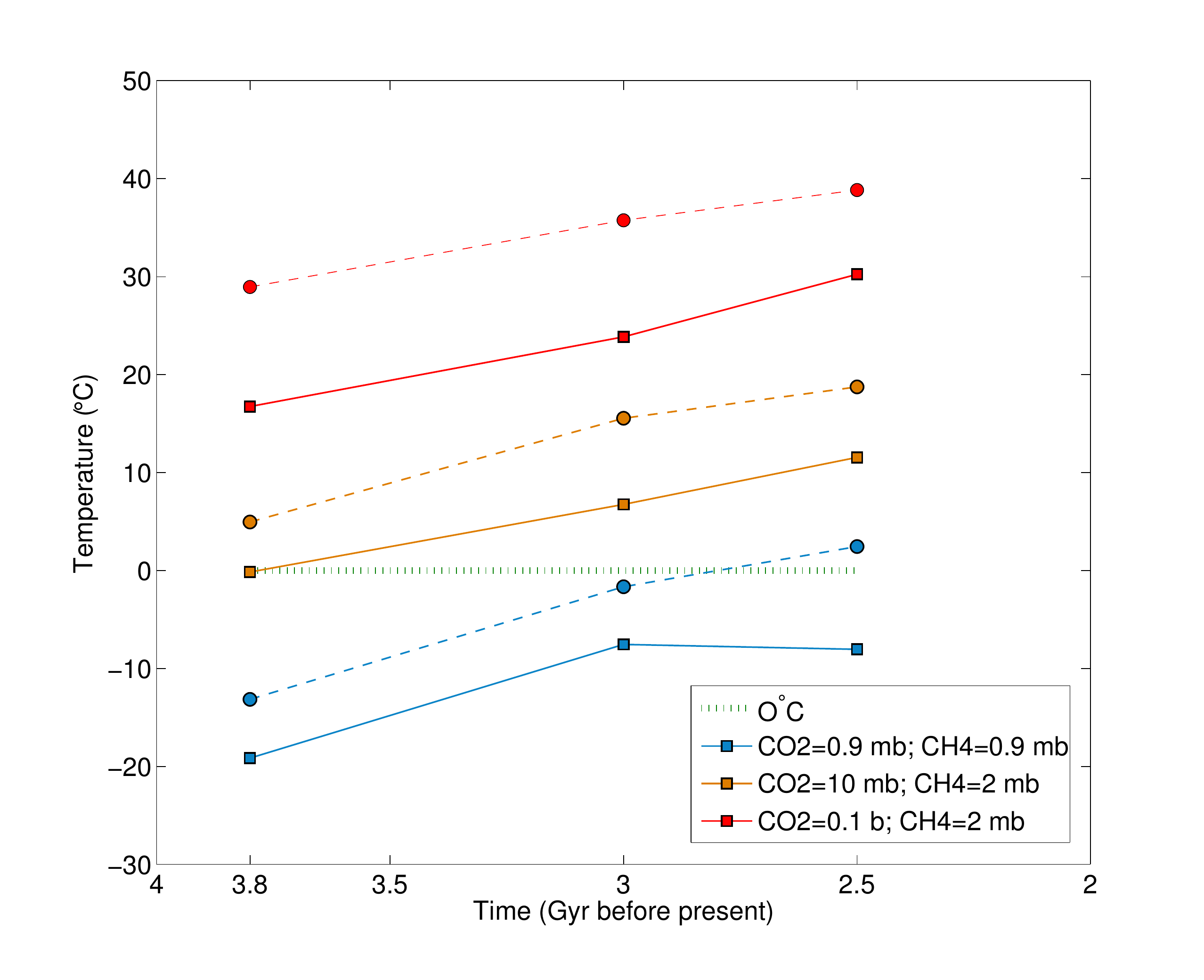}
\caption{Same as Figure 7 with liquid droplet radius of 12 $\mu$m (fill lines) and 17 $\mu$m (dashed lines).}
\label{figure12}
\end{figure}

Here, we compare simulations with the radius for liquid clouds fixed at 17 $\mu$m to simulations with the radius fixed at 12 $\mu$m (present-day value used before). 
Figure \ref{figure11} shows the mixing ratio of water condensed in the atmosphere (i.e. the mass of cloud per mass of air) at 2.5 Ga with the atmospheric composition B for both radii, and the difference between both. There are less lower clouds (i.e. lower than 5 km) for low and mid latitudes in the case of 17 $\mu$m due to the increase in precipitation rate. There are more lower clouds at high latitudes and more higher clouds (i.e. higher than 5 km) at low and mid latitudes. This is due to the warmer surface and to the Hadley cell, which extends deeper in the troposphere. We find a reduction by 10 to 20 $\%$ of the amount of low and mid latitude lower clouds, in agreement with equation (\ref{precip}) to maintain the same precipitation rate. 
In the tropics (i.e. latitudes between 30$^\circ$N and 30$^\circ$S), the mean optical depth of clouds decreases by around 30$\%$ (from 37.2 to 26.9.).
Because of the larger droplets and the lower amount of lower clouds, the net solar absorption increases by 10.3 W/m$^2$ and the planetary albedo changes from 0.33 to 0.29. The mean surface temperature changes from 11.6 $^\circ$C to 18.8 $^\circ$C with the larger radius, which corresponds to a warming of +7.2$^\circ$C. Around 4.8$^\circ$C of warming are due to the direct effect of larger droplets on optical depth (Twomey effect) and around 2.4$^\circ$C of warming are due to the enhanced precipitations (Albrecht effect). We conclude that the first effect is dominant with our model (see next paragraph).
Fig. \ref{figure12} presents the mean surface temperature throughout the Archean for our 3 different atmospheric compositions and with radii of 12 and 17 $\mu$m. The warming for every case due to the larger droplets is significant, between +5$^\circ$C and +12$^\circ$C.

\begin{figure}[h!]
\centering
\noindent\includegraphics[width=20pc]{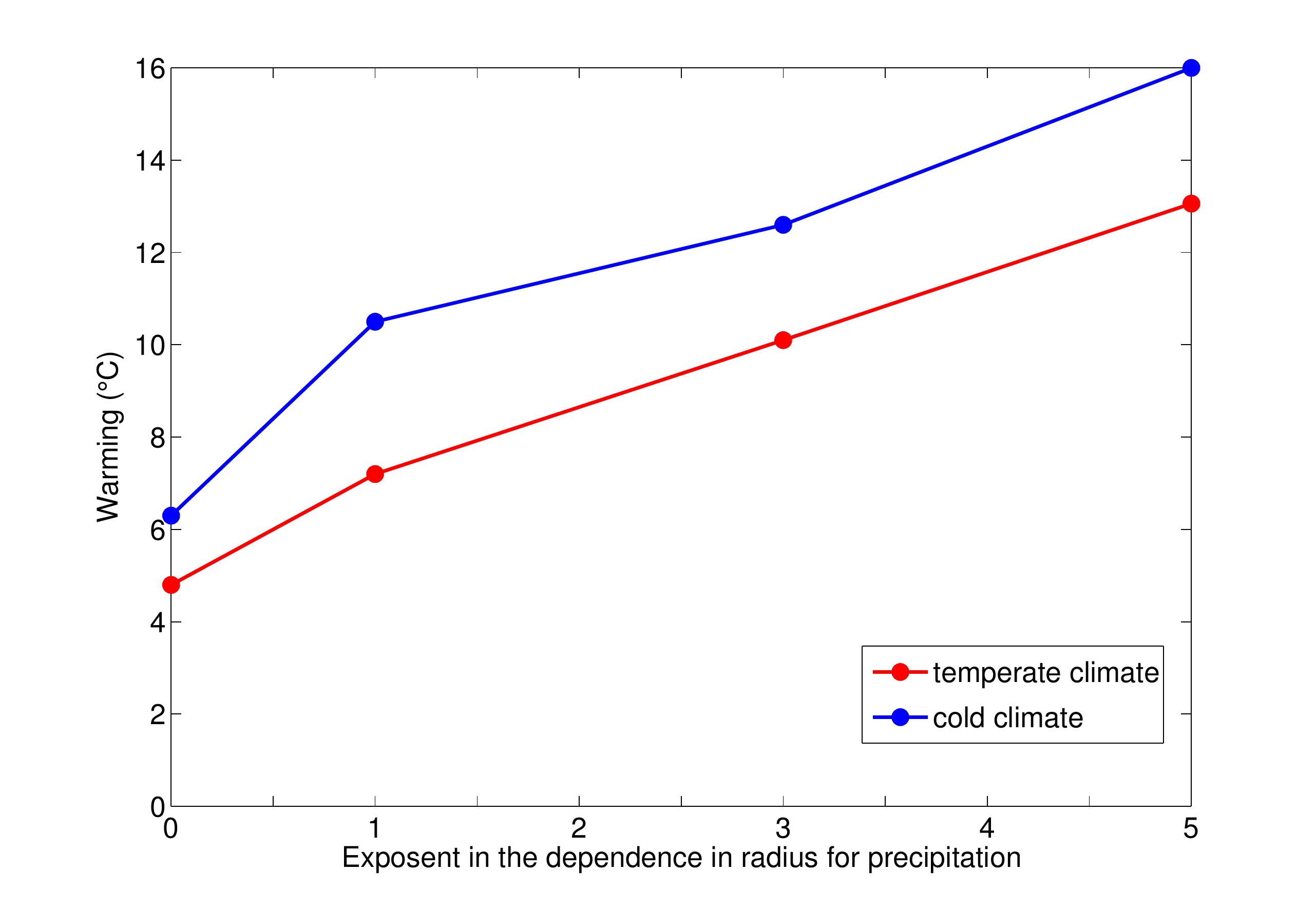}
\caption{Warming produced by the increase of liquid droplet radius from 12 to 17 $\mu$m as a function of the exponent n for the assumption that the precipitation rate is proportionnal to (r/r$_0$)$^n$. Simulations at 2.5 Ga with 10 mb of CO$_2$ and  2 mb of CH$_4$ (red), and 0.9 mb of CO$_2$ and CH$_4$ (blue).}
\label{figure13}
\end{figure}

We now investigate the impact of the change in the exponent on the dependence of radius in the precipitation rate.
Figure \ref{figure13} shows the mean surface temperature for the late Archean with the atmospheric composition B, radius of 17 $\mu$m and the exponent varying from 0 to 5 for cold (atmospheric composition A) and temperate climates (atmospheric composition B). The warming is stronger starting from a cold climate owing to the higher climate sensitivity linked to the ice-albedo feedback under such conditions. Even with no change of the precipitation rate (exponent 0), the warming is significant (between +4.5$^\circ$C and +6$^\circ$C). The warming obtained with the exponent 3 ($\sim$+10$^\circ$C) is consistent with the model from \cite{rosing10}. Yet, even with the highest exponent, which still remains rather unrealistic, we cannot reach present-day temperatures with the atmospheric composition A. This analysis also reveals that the Twomey effect is dominant for exponents lower than 3, and the Albrecht effect is dominant for exponents higher than 3.

To conclude, larger cloud droplets produce a very efficient warming, even if a large uncertainty remains on the impact of cloud precipitation. That could have helped the Archean Earth to reach a temperate or warm climate but it is not sufficient, according to our model, to solve the faint young Sun problem alone. A significant amount of greenhouse gases is still required. Moreover, the possibility to have droplets reaching 17 $\mu$m is still speculative.

\subsection{Effect of the atmospheric pressure}
With the lack of oxygen, the N$_2$ mixing ratio was probably larger in the Archean than it is presently, making nitrogen the main gas governing atmospheric pressure. Nitrogen is an inert gas with a negligible direct greenhouse effect in terrestrial conditions. However, an increase in atmospheric pressure reinforces the effect of other greenhouse gases through pressure broadening. In the troposphere, the temperature profile is globally often determined by the moist adiabatic lapse rate corresponding to the stability limit of air against moist convection. The moist adiabatic lapse rate given by \cite[]{sanchez-lavega_book}

\begin{equation}
\Gamma_s=-\left(\frac{dT}{dz}\right)_S = \frac{g}{c_p}\left(\frac{1+\frac{L_{V} x_S}{R_dT}}{1+\frac{L_{V}^2 x_S}{c_p R_d T^2}}\right)
\end{equation}
where g is the gravity, $c_p$ is the heat capacity of dry air, $L_{V}$ is the latent heat of vaporization of water, $T$ is the temperature, $R_d$ and $R_v$ are the gas constants for dry air and water vapor  $x_S=\epsilon\frac{P_{VS}(T)}{P}$ is the saturation mixing ratio where $P_{VS}(T)$ is the saturation vapor pressure, $P$ is the pressure and $\epsilon=M_w/M_d$ is the ratio of molar masses of water and dry air.
An increase in the atmospheric pressure necessarily increases the moist adiabatic lapse rate (going closer to the dry adiabatic lapse rate $\frac{g}{c_p}$=-9.8 K/km). This leads to an additional warming. In return, an increased pressure leads to enhanced Rayleigh scattering by the atmosphere, increasing the planetary albedo, and cooling the surface.

It has been suggested that the pressure was higher in the past \cite[]{goldblatt09}. The equivalent of about 2 bars of nitrogen is present in the Earth's mantle. That nitrogen 
was likely initially present in the atmosphere and later fixed by surface organims and incorporated into the mantle by subduction. This is corroborated by the correlation between N$_{2}$ and radiogenic $^{40}$Ar contrary to primordial $^{36}$Ar in bubbles in oceanic basalt \cite[]{goldblatt09, marty03}. Nitrogen fixation is biological and would have increased strongly with the appearance of photosynthetic life. That process would have led to a decrease of the amount of nitrogen in the atmosphere \cite[]{goldblatt09}.

Thus the partial pressure of nitrogen may have reached 2 to 3 bars during the Archean. According to 1D modelling, a doubling of the present-day atmospheric nitrogen amount could cause a warming of 4-5 $^\circ$C \cite[]{goldblatt09}.
The change in the moist adiabtic lapse rate impacts the formation of convective clouds, and hence the quantity of clouds in the atmosphere. Moreover, a higher atmospheric pressure impacts the transport of heat to poles. Neither are taken into account in 1D models. 

\begin{figure}[h!]
\centering
\includegraphics[width=20pc]{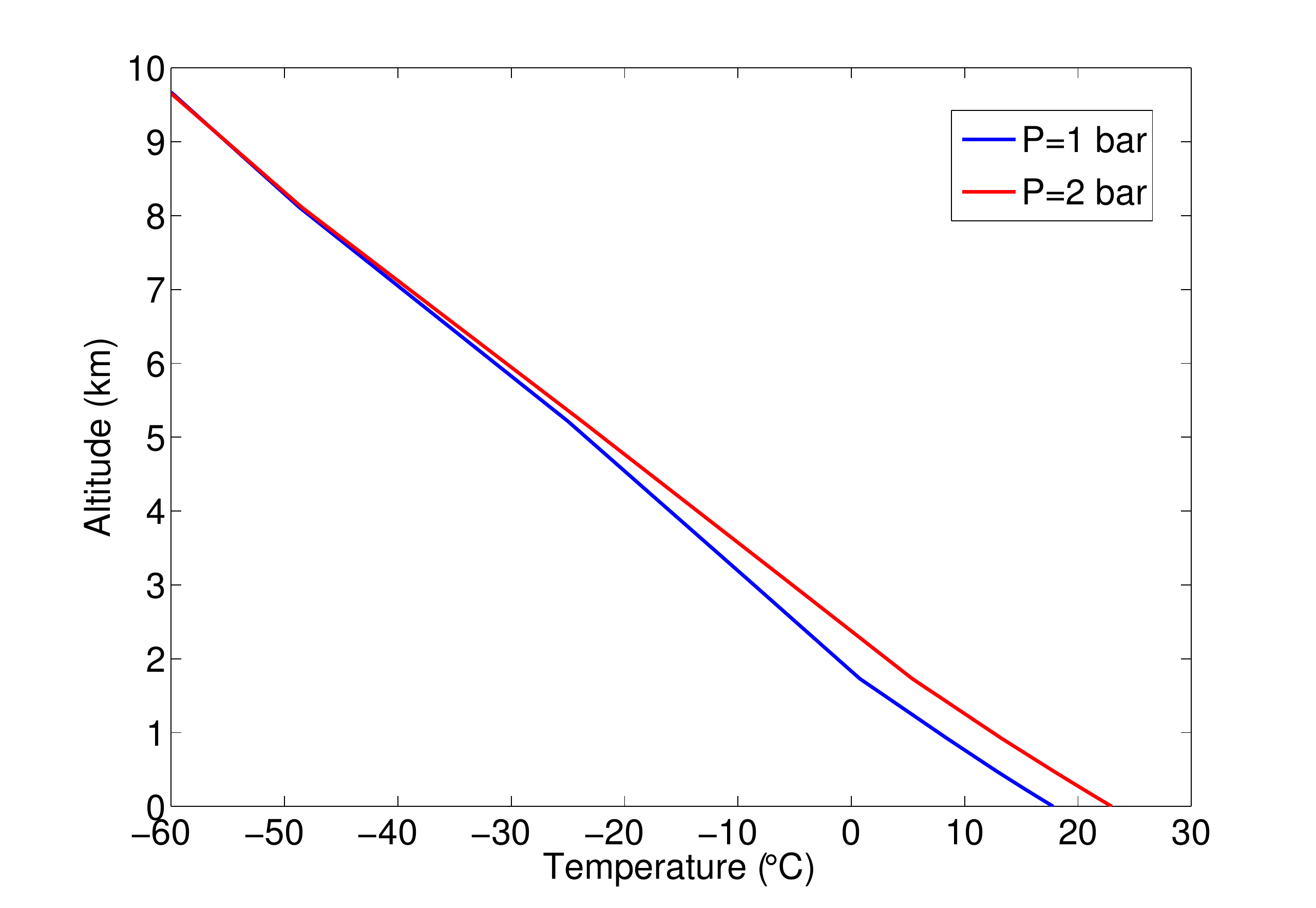}\\
\includegraphics[width=20pc]{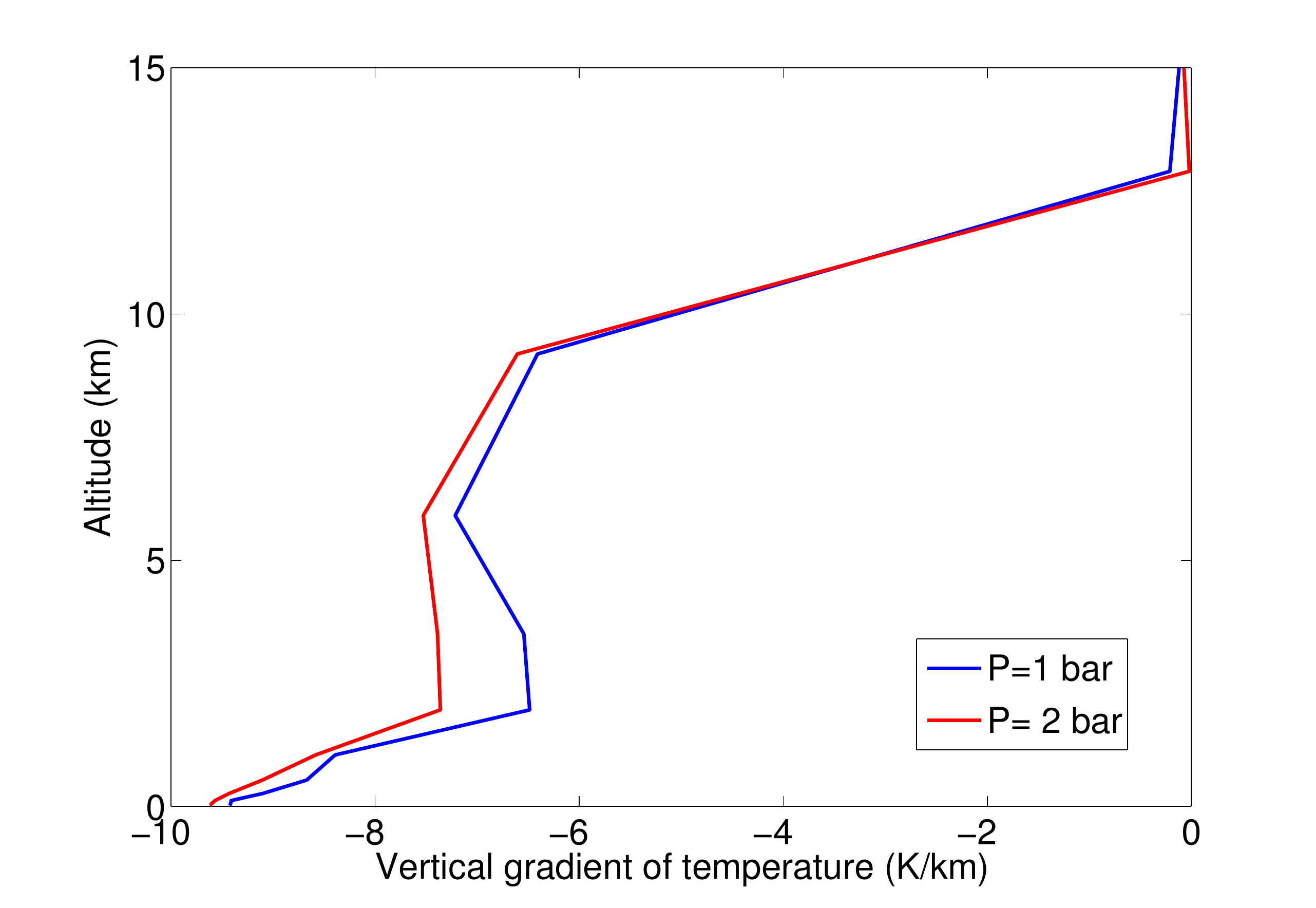}
\caption{Zonally averaged vertical temperature profiles (top) and temperature lapse rate (down) at the equator at 2.5 Ga (atmospheric composition: 10 mb of CO$_2$ and  2 mb of CH$_4$) with an atmospheric pressure of 1 bar (blue) and 2 bars (red).}
\label{figure14}
\end{figure}

We have run a simulation for the end of the Archean with the atmospheric composition B and 2 bars of nitrogen.
Figure \ref{figure14} illustrates the change in the atmospheric lapse rate. In the troposphere, the temperature lapse rate increases when pressure is doubled. However, we do not notice particular changes in clouds. With the atmospheric composition B we get a warming of +7$^\circ$C so 2-3$^\circ$C more than the warming obtained by \cite{goldblatt09}. The difference is probably related to the higher amount of CH$_4$ in our simulation (2 mb versus 0.1 mb). 
The accumulated effects of larger droplets and doubling pressure produce a warming of around 10.5$^\circ$C, less than the sum of both effects taken separately ($\sim$14$^\circ$C).

A higher surface pressure produces an efficient warming of the same magnitude as larger droplets. Yet, an analysis of raindrop impact sizes suggest that the air density at 2.7 Ga was close to the modern value and limited to less than twice the modern value \cite[]{som12}. Thus, the surface pressure was probably close to 0.8-1 bar at the end of the Archean, even if up to 2 bars remain plausible. The pressure may have been higher at older periods, in particular at the beginning of the Archean. Yet recent data of nitrogen isotopes at 3.5 Ga \cite[]{marty12} seems not go that way. This implies questions about when was the nitrogen incorporated into the mantle and by what means.

\subsection{Effect of the rotation rate}
The Moon's tides produce friction which tends to slow down Earth's rotation. By conservation of angular momentum, the Moon tends to move away from Earth \cite[]{williams00, walker86}. Thus, the Earth was rotating faster in the past. Earth's rotation period is estimated to be around 14 hours at 4 Ga \cite[]{zahnle87}. A faster rotation rate impacts the climate by reducing the meridional transport \cite[]{stone72, hunt79, kuhn89, feulner12}. The meridional transport at mid-latitudes is mainly produced by baroclinic eddies, whose size is scaled to the Rossby deformation radius $L_D$, which is linked to the rotation rate $\Omega$ by \cite[]{sanchez-lavega_book}

\begin{equation}
L_D=\frac{N_B H}{f_0}
\end{equation}
where $N_B$ is the Brunt-V\"ais\"al\"a frequency, $H$ is the scale height and $f=2 \Omega sin \phi$ (with $\phi$ the latitude) is the Coriolis parameter.

A faster rotation rate limits the size of eddies, which decreases the efficiency of meridional transport. Under such conditions, the equator-pole thermal gradient is enhanced with a warmer equator and cooler poles. 

\begin{figure*}[h!]
\centering
\noindent\includegraphics[width=20pc]{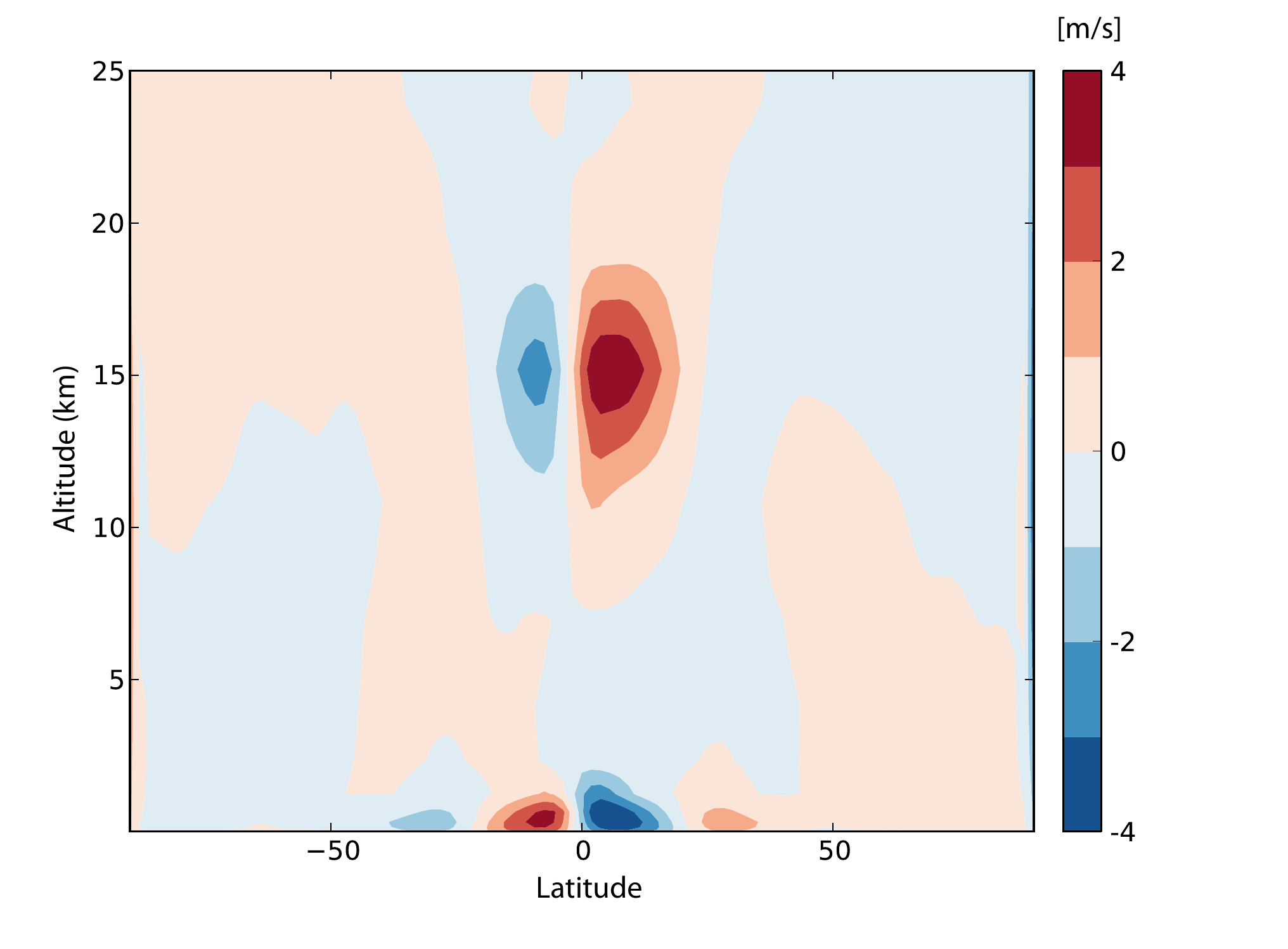}
\noindent\includegraphics[width=20pc]{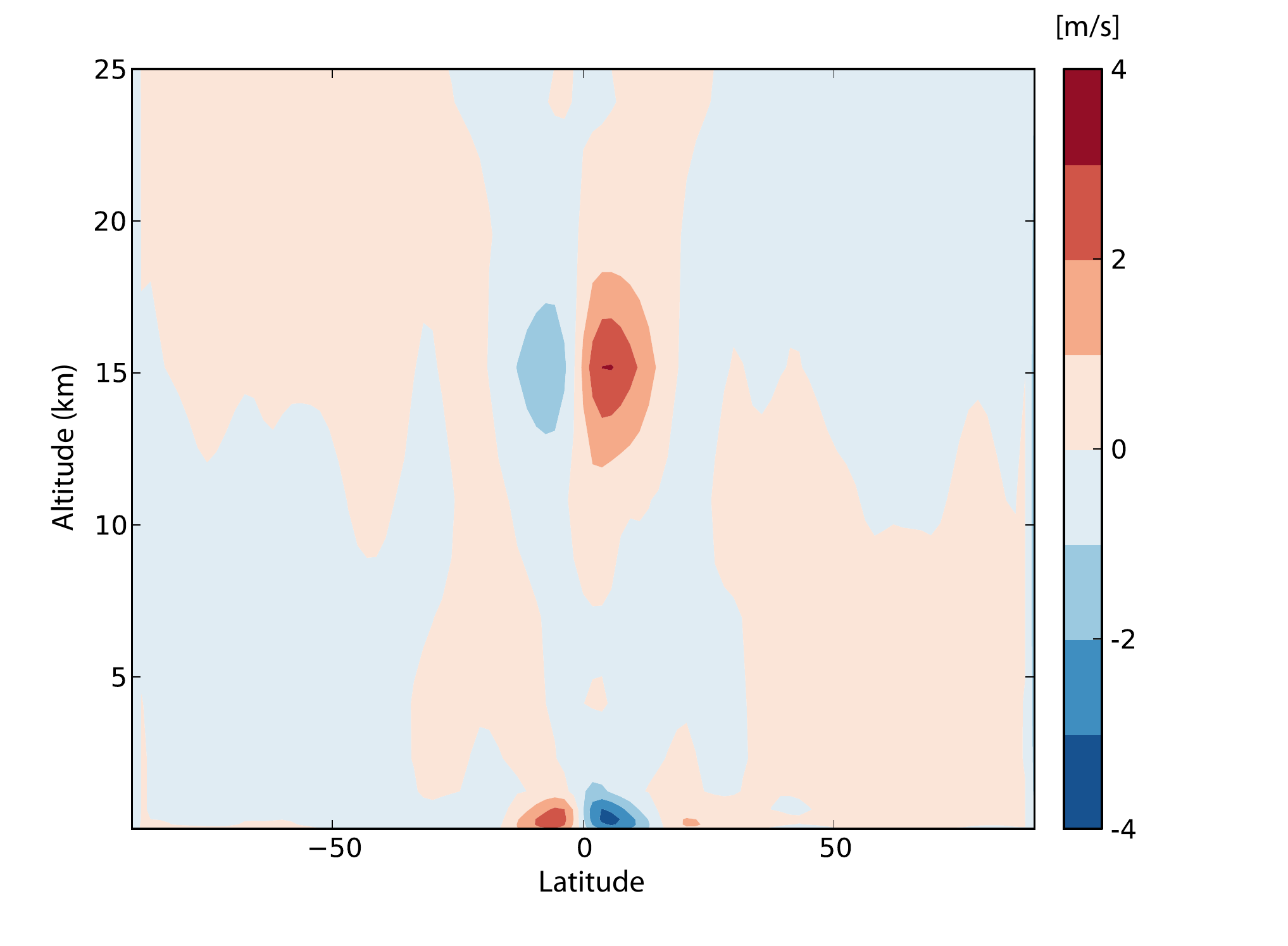}
\noindent\includegraphics[width=20pc]{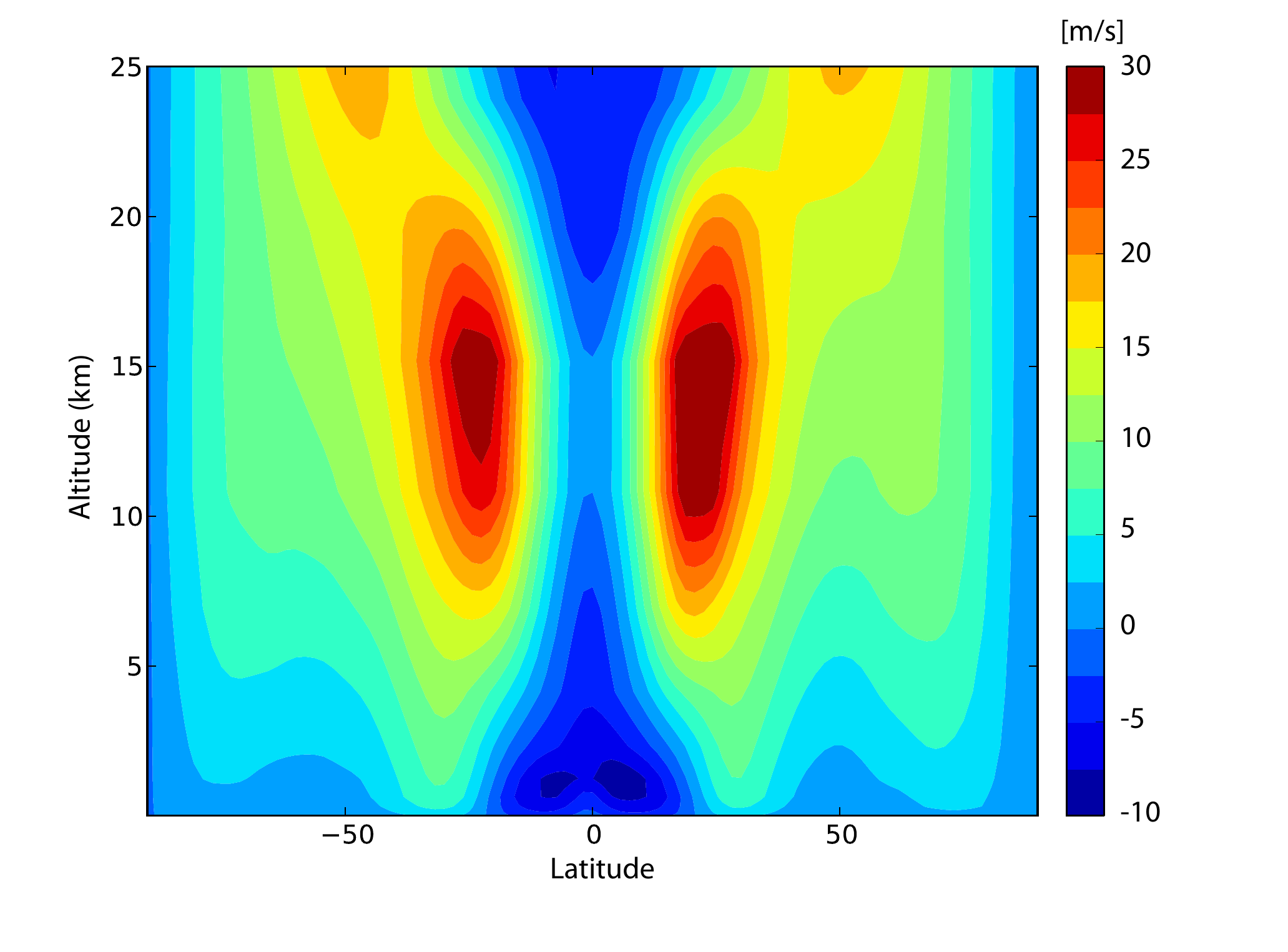}
\noindent\includegraphics[width=20pc]{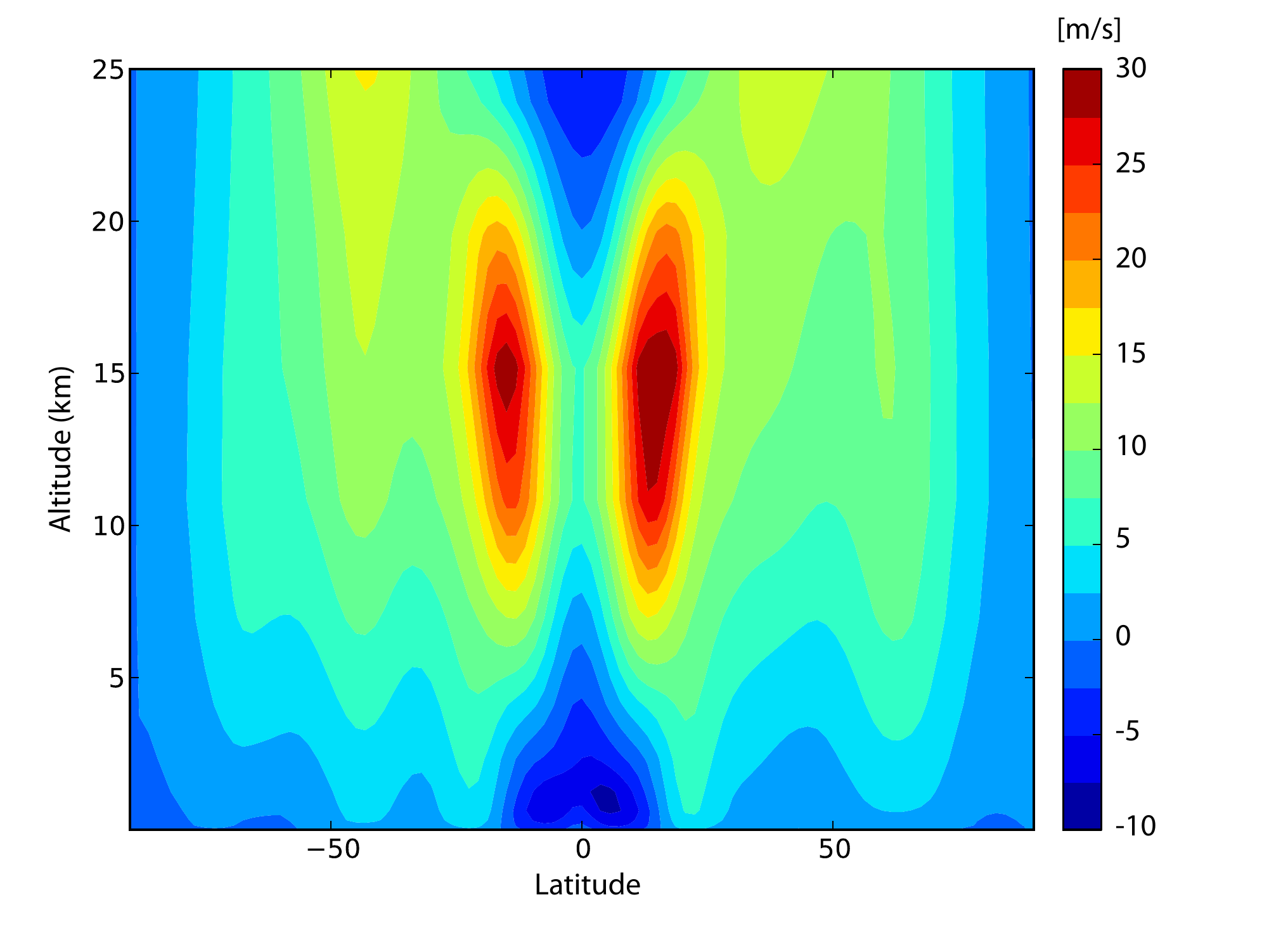}
\caption{Zonally averaged meridional (top) and zonal (bottom) winds for a rotation period of 24 hours (left) or 14 hours (right). Simulations at 3.8 Ga with 10 mb of CO$_2$, 2 mb of CH$_4$ and droplet radius of 17 $\mu$m.}
\label{figure15}
\end{figure*}

We ran the model at 3.8 Ga with a rotation rate of 14 hours for different conditions: a temperate, a warm and a cold climate. For the temperate climate (10 mb of CO$_2$, 2 mb of CH$_4$ and droplet radius of 17 microns), we found a limited impact of the faster rotation rate. The mean surface temperature in the temperate climate is 1$^\circ$C lower with the faster rotation rate, and sea ice is more extended, but no significant change in cloud covering and planetary albedo is observed. Figure \ref{figure15} shows the winds for both rotation rates for the temperate climate. With the faster rotation rate, the latitudinal extension and the strength of the Hadley cells is reduced, and the jets are closer to the equator. 

\begin{figure*}
\centering
\includegraphics[width=20pc]{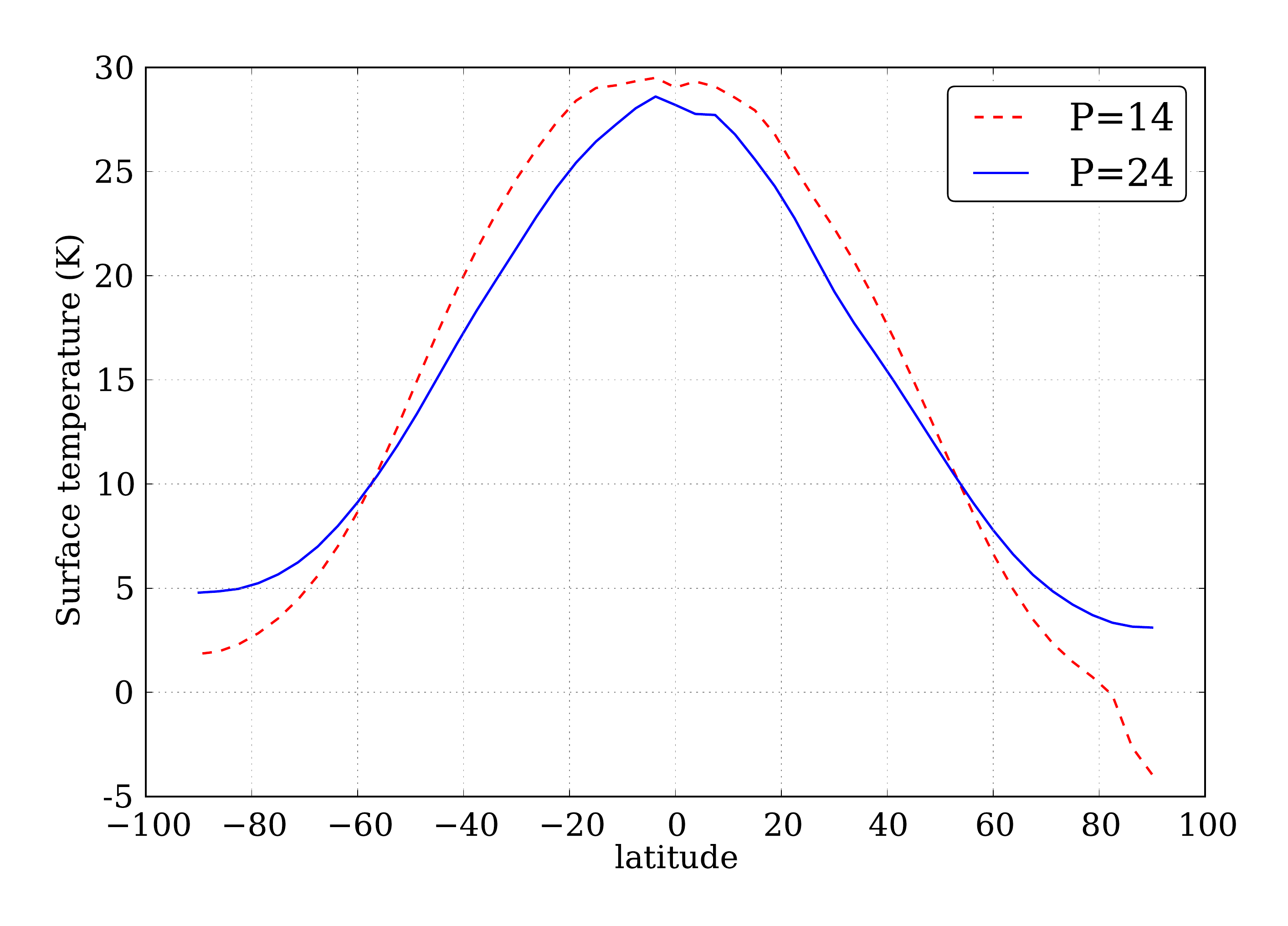}
\includegraphics[width=20pc]{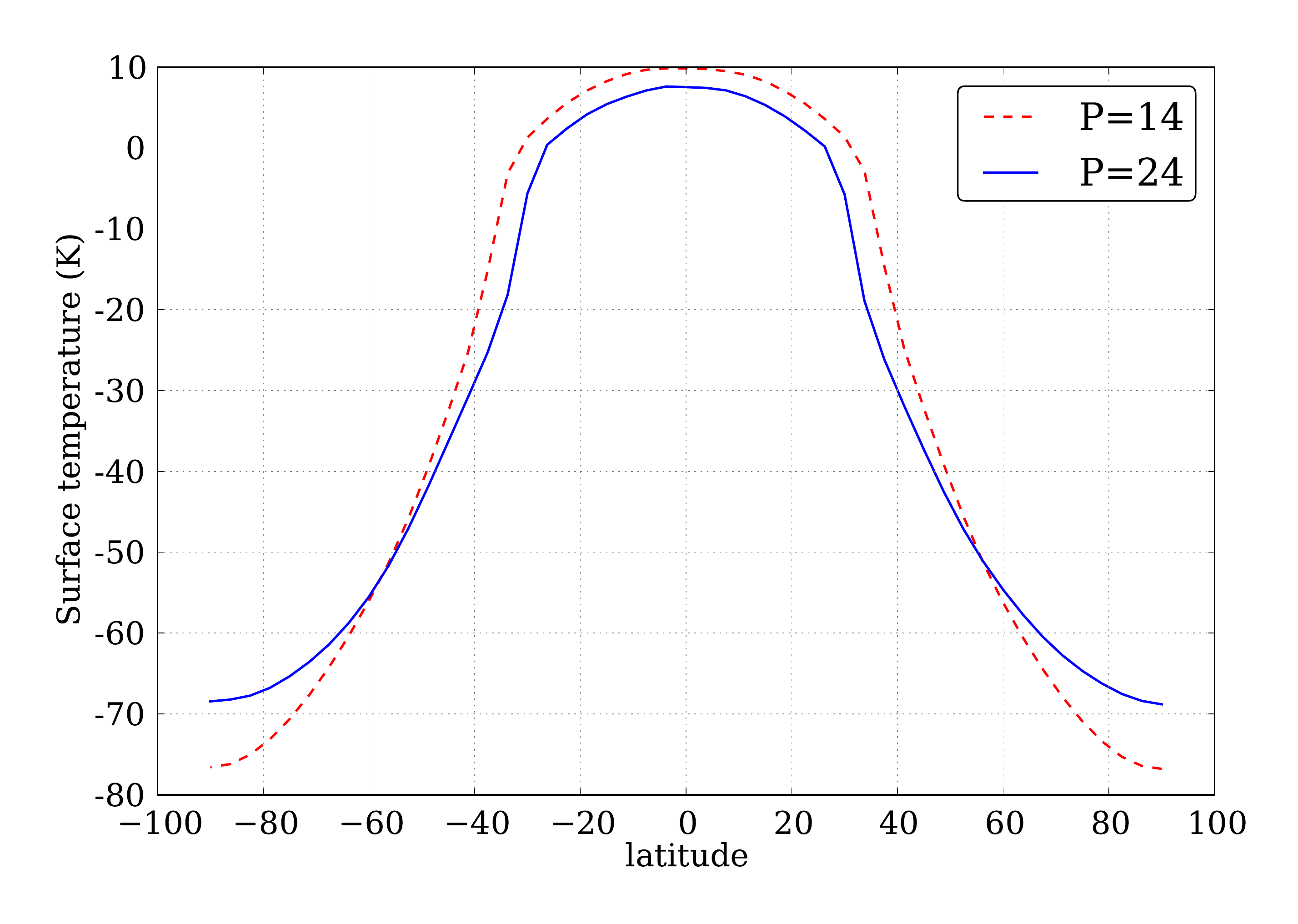}
\caption{Zonally-averaged surface temperature at 3.8 Ga with 100 mb of CO$_2$, 2 mb of CH$_4$ (left), and 0.9 mb of CO$_2$ and CH$_4$ (right). The rotation period is either 14 hours (red dashed line) or 24 hours (blue fill line).}
\label{figure16}
\end{figure*}

In contrast, for a cold and a warm climate (climate with no oceanic ice), a faster rotation rate produces a warming with our model.
Figure \ref{figure16} shows the surface temperature at 3.8 Ga for a warm and a cold climate with a rotation rate of either 24 or 14 hours.  The mean surface temperature in the warm climate is 1.5$^\circ$C higher with the faster rotation rate while the mean tropical oceanic temperature is around 2$^\circ$C higher. Since no ice-albedo happens in this case, the mean temperature increases due to the water vapor feedback, which is enhanced in the tropics. 
In the case of a cold climate, the faster rotation and the reduced meridional transport limit the propagation of sea ice to the equator since the removal of heat from the tropics is less efficient. Waterbelts are larger (Figure \ref{figure16}) and the triggering of full glaciation is weakened. To summarize, the rotation rate seems to have a very limited impact on the climate of the early Earth. The only notable impact is to make a full glaciation a little bit more difficult.

\section{Summary and discussion}
In this paper, we have tested a range of hypotheses to solve the faint young Sun problem and we have obtained new estimates for the greenhouse gases required for a temperate climate for the Archean Earth.
We have shown that a CO$_2$-CH$_4$ rich atmosphere is a viable solution to the faint young Sun problem. A composition of 10 mb of CO$_2$ with 2 mb of CH$_4$ allows us to get a temperate climate (mean surface temperature between 10$^\circ$C and 14$^\circ$C, so just a little colder than today) at the end of the Archean while satisfying the geological constraints on the CO$_2$ partial pressure. At the beginning of the Archean, a partial pressure of around 0.1 bar of CO$_2$ with 2 mb of CH$_4$ is required to get temperate climates if no other warming mechanism occurred. However, no geological constraints for the CO$_2$ partial pressure at this period currently exist. The presence of only CO$_2$ would not provide a sufficient greenhouse warming to get temperate climates. The removal of methane at 2.5 Ga produced by the oxidation of the atmosphere in the early Proterozoic yields a cooling of 14$^\circ$C. This strong glaciation is not global according to our model. This could be a good scenario for the glaciations of the early Proterozoic and preserving life in the ice-free waterbelt. 

The Archean Earth most likely featured less emerged land than today. However a change in land distribution has only a small influence on the ocean temperatures (less than 3$^\circ$C over the whole Earth and less than 1.5$^\circ$C over the tropics). The lack of evidence of glaciation during most of the Archean can be interpreted as being caused by a hotter climate than today but also as another distribution of land for a similar or colder climate. An Earth with little emerged land that is essentially concentrated at the equator will only keep geological evidence of full glaciations.
The faster rotation of the Archean Earth seems to have a very small impact on the global climate. However, it can limit the triggering of a full snowball Earth a little.
Larger cloud droplets appear to be a very efficient way to warm the Archean earth. The uncertainty in the precipitation rate implies a large uncertainty in the warming, but it remains strong in any case in our simulations (warming between +5$^\circ$C and +13$^\circ$C). A better understanding of the different processes and feedbacks implied in precipitation will be required to refine this effect. Moreover, a reduced number of CCN could impact the formation of higher clouds. This could amplify or diminish the warming.
However, larger droplets during the Archean have yet to be confirmed through geological data, which will be challenging. 
A higher atmospheric pressure also appears to be an efficient warming process. However, analyses of raindrops imprints \cite[]{som12} along with recent data \cite[]{marty12} challenge this process for most of the Archean.
These last two hypotheses are not sufficient to solve the faint young Sun problem but they may have played a significant role in complementing the greenhouse effect to CO$_2$ and CH$_4$. Table 3 summarizes the different warming processes at 2.5 Ga with 10 mb of CO$_2$ with 2 mb of CH$_4$.

Our GCM supports some of the main conclusions of 1D models, but it also reveals some interesting 3D behavior of the climate. There is a decrease in clouds above continents that compensates for their higher surface albedo and there is especially an important cloud feedback. For the same surface temperature, there are less clouds during the Archean due to the weaker evaporation. This leads to a lower planetary albedo. In our model, another strong cloud feedback appears in cold climates. The decrease of clouds particularly close to the freezing line counteracts the ice-albedo feedback and allows waterbelts to exist with a mean surface temperature far below the frozen point. Such a resistance against glaciation mitigates the faint young Sun problem.

\section{Perspectives}
This article constitutes the first full study of the Archean Earth with a 3D GCM coupled to a dynamic oceanic model. 
Coupled 3D models applied to the early Earth provide new tools for achieving progress in a field where there is still a lot to understand. 3D GCMs have to be as general (less tuned) as possible. As for global warming, comparison between several GCMs will be required to refine predictions.
Our study consolidates many results obtained with 1D radiative-convective models and emphasizes some particular behaviors inherent to 3D as well. There are many ways to solve the faint young Sun problem. According to our results, it is not so difficult a task, particularly if cold climates with waterbelts can be maintained. New geological constraints for the early Archean on CO$_2$ and N$_2$ partial pressures, as well as the H$_2$ abundance of the atmosphere, are necessary to have a good picture of the atmospheric composition.

There are many perspectives for future research with our new modelling tool. First, the levels of CH$_4$ we used could lead to the formation of organic haze, in particular for the end of the Archean, where some geological data could be interpreted as the episodic formation of haze \cite[]{zerkle12}. The study of the formation, dynamics and anti-greenhouse effect of organic haze with a 3D model would constitute major progress on this topic. Finally, although it appears possible to get a temperate climate during the Archean, producing a hot early Earth with oceans at 60$^\circ$C to 80$^\circ$C, as suggested from oceanic cherts \cite[]{robert06}, requires a far stronger warming. Although this idea is controversial, it would be interesting to see what pressure of CO$_2$, in addition to other warming processes (higher pressure, larger cloud droplets, methane) would be required, more precisely than with 1D models. Such a study would also be applicable to the Hadean Earth.

%%% End of body of article:

%%%%%%%%%%%%%%%%%%%%%%%%%%%%%%%%
%% Optional Appendix goes here
%
% \appendix resets counters and redefines section heads
% but doesn't print anything.
% After typing  \appendix
%
% \section{Here Is Appendix Title}
% will show
% Appendix A: Here Is Appendix Title
%
%\appendix
%\section{Details on the mathematical formulation of the thermal plume model}

%%%%%%%%%%%%%%%%%%%%%%%%%%%%%%%%%%%%%%%%%%%%%%%%%%%%%%%%%%%%%%%%
%
% Optional Glossary or Notation section, goes here
%
%%%%%%%%%%%%%%
% Glossary is only allowed in Reviews of Geophysics
% \section*{Glossary}
% \paragraph{Term}
% Term Definition here
%
%%%%%%%%%%%%%%
% Notation -- End each entry with a period.
% \begin{notation}
% Term & definition.\\
% Second term & second definition.\\
% \end{notation}
%%%%%%%%%%%%%%%%%%%%%%%%%%%%%%%%%%%%%%%%%%%%%%%%%%%%%%%%%%%%%%%%
%
%  ACKNOWLEDGMENTS

\begin{acknowledgments}
We thank Olivier Boucher for discussions of cloud formation and precipitations. 

\end{acknowledgments}

\end{article}

%% Enter Figures and Tables here:

\clearpage

\begin{table}
\begin{tabular}{|m{5.1cm}|m{2.4cm}|m{3cm}|m{1.8cm}|m{2.2cm}|m{2.2cm}|}
\hline
   Land distribution&Mean surface temperature &Tropical oceanic temperature& Planetary albedo & Precipitable water & Condensed water \\ \hline 
   Modern Earth (present Sun) &14.8$^\circ$ C & 23.7$^\circ$ C& 0.36 & 18.15 kg/m$^2$ & 0.16 kg/m$^2$ \\
   Present-day land & 10.5$^\circ$ C  & 20.0$^\circ$ C& 0.34&14 kg/m$^2$ & 0.135 kg/m$^2$ \\
   Equatorial supercontinent & 11.6$^\circ$ C & 19.4$^\circ$ C& 0.33 & 13.2 kg/m$^2$ & 0.12 kg/m$^2$\\
   No land & 13.8$^\circ$ C & 20.8$^\circ$ C& 0.33 & 16 kg/m$^2$ & 0.15 kg/m$^2$ \\ \hline
\end{tabular}
\label{table1}
\caption{}
\end{table}

\begin{table}
\begin{tabular}{|m{5cm}|m{2.4cm}|m{1.9cm}|m{2.3cm}|m{3cm}|}
\hline
Case						&	Mean surface temperature	&	Planetary albedo  	&	Condensed water	&	Precipitation 		\\ \hline
Aqua-planet (present Sun)	&	19$^\circ$ C 				&	0.36 			&	0.19 kg/m$^2$		&	4.0e-5 kg/m$^2$/s 	\\
Earth at 3.8 Ga				&	18.7$^\circ$ C 				&	0.30		 		&	0.16 kg/m$^2$		&	3.4e-5 kg/m$^2$/s 	\\
\hline
\end{tabular}
\label{table2}
\caption{}
\end{table}

\begin{table}
\begin{tabular}{|l|c|}
\hline
   Warming process&Warming \\ \hline
   Methane (2 mb) & +14$^\circ$ C \\
   Less land & $\le$ +3.5$^\circ$ C \\
   Larger droplet & +7$^\circ$ C \\
   Doubling pressure & +7$^\circ$ C \\
   Larger droplet + Doubling pressure & +10.5$^\circ$ C \\
   Faster rotation rate & -1$^\circ$ C \\
\hline
\end{tabular}
\label{table3}
\caption{}
\end{table}

\end{document}